\documentclass[11pt,twoside,letterpaper]{article} 
\usepackage{times,fancyhdr}
\usepackage[dvips]{graphicx}
\pdfoutput=1
\makeatletter 
\def\cleardoublepage{\clearpage\if@twoside \ifodd\c@page\else%
    \hbox{}%
    \thispagestyle{empty}%
    \newpage%
    \if@twocolumn\hbox{}\newpage\fi\fi\fi} 
\makeatother 
\def\figurename{Figure}
\makeatletter
\renewcommand{\fnum@figure}[1]{\figurename~\thefigure.}
\makeatother
\def\tablename{Table}
\makeatletter
\renewcommand{\fnum@table}[1]{\tablename~\thetable.}
\makeatother
\usepackage{amssymb}
\usepackage{amsmath}
\usepackage{amsfonts}
\usepackage{amsthm,amscd}
\usepackage{amsbsy}
\usepackage{latexsym}
\usepackage{bm}
\usepackage{url} 			
\usepackage{layout}
\usepackage{pslatex}
\usepackage{fleqn} 		
\usepackage{makeidx}
\makeindex 						
\usepackage{layout} 	
\usepackage{epstopdf}	
\usepackage{color}
\usepackage[latin1]{inputenc}
\usepackage[T1]{fontenc}
\usepackage{poligraf} 
\usepackage[letter,cam,center]{crop} 
\usepackage{type1cm} 	
\usepackage{courier}	
\usepackage{lscape} 	
\newcommand{\lSect}[1]{{\label{sec:#1}}}
\newcommand{\Sectff}[1]{{\ref{sec:#1}}}
\newcommand{\Sect}[1]{{\S\Sectff{#1}}}

\newcommand{\be}{\begin{eqnarray}} 
\newcommand{\ee}{\end{eqnarray}}

\newcommand\gsim{\buildrel > \over \sim}
\newcommand\lsim{\buildrel < \over \sim}

\newcommand\simless{\buildrel < \over \sim}

\begin{document}
\title{
{\begin{flushleft}
\vskip -0.5in
{\normalsize INT-PUB-12-002}
\end{flushleft}
\vskip 0.45in
\bfseries\scshape Thermal and transport properties of the neutron star inner crust}}

\author{\bfseries\itshape Dany Page\thanks{E-mail address:page@astro.unam.mx},  Sanjay Reddy \thanks{E-mail address:sareddy@uw.edu}\\
Instituto de Astronom\'ia, Universidad Nacional Auton\'oma de M\'exico, \\ Mexico D.F. 04510, Mexico \\
Institute for Nuclear Theory, University of Washington, Seattle, Washington 98195 \\
}
\date{}
\maketitle

\abstractname{: 
We review the nuclear and condensed matter physics underlying the thermal and transport properties of the neutron star inner crust.  These properties play a key role in interpreting transient phenomena such as thermal relaxation in accreting neutron stars, superbursts, and magnetar flares. We emphasize simplifications that occur at low temperature where the inner crust can be described in terms of electrons and collective excitations. The heat conductivity and heat capacity of the solid and superfluid phase of matter is discussed in detail and we emphasize its role in interpreting observations of neutron stars in soft X-ray transients. We highlight recent theoretical and observational results, and identify future work needed to better understand a host of transient phenomena in neutron stars.   
}

\thispagestyle{empty}
\setcounter{page}{1}
\thispagestyle{fancy}
\fancyhead{}
\fancyhead[L]{In: Neutron Star Crust \\ 
Editor:  C. A. Bertulani  \& J. Piekarewicz {\thepage-\pageref{lastpage-01}}} 
\fancyhead[R]{ISBN 0000000000  \\
\copyright~2007 Nova Science Publishers, Inc.}
\fancyfoot{}
\renewcommand{\headrulewidth}{0pt}

\vspace{0.2in}

\vspace{.08in} 


\pagestyle{fancy}
\fancyhead{}
\fancyhead[EC]{Page \& Reddy}
\fancyhead[EL,OR]{\thepage}
\fancyhead[OC]{}
\fancyfoot{}
\renewcommand\headrulewidth{0.5pt} 

\section{Introduction} 

Left undisturbed, neutron stars quietly fade away on a time-scale of about a million years as neutrinos from the core and, later, photons from the surface sap their thermal energy. The neutron star core, mysterious as it still remains, is expected to be a good conductor of heat, and with sufficient mass and heat capacity it acts as an excellent heat reservoir. In contrast, the crust is composed of matter that can be activated by external perturbations such as accreting matter, evolving magnetic fields and angular torques. Several observed transient phenomena such as X-ray bursts, superbursts, magnetar flares and glitches in the spin evolution are thought to originate in the crust. X-ray bursts occur at relatively shallow depth, at low density and high temperature, while superbursts, flares, glitches and heating during accretion are expected to occur deeper in the crust, in denser and more degenerate conditions. In these latter phenomena, thermal and transport properties differ from those expected in ideal gases because of strong Coulomb and nuclear interactions, matter degeneracy, and superfluidity in the inner crust. Interestingly, these properties play an important role in shaping temporal aspects of the observed X-ray light-curves. Recent observations of transient phenomena in accreting and magnetized neutron stars have renewed interest in developing realistic models of the neutron star crust.

In this chapter we discuss the basic physics of the inner crust from the viewpoint of a low energy theory of electrons and phonons. For a more comprehensive survey we refer the reader to an excellent review article by Chamel and Haensel \cite{Chamel:2008LRR}. 
\section{The composition of the inner crust}
\begin{figure}[htbp]
   \centering
   \includegraphics[width=0.7\textwidth]{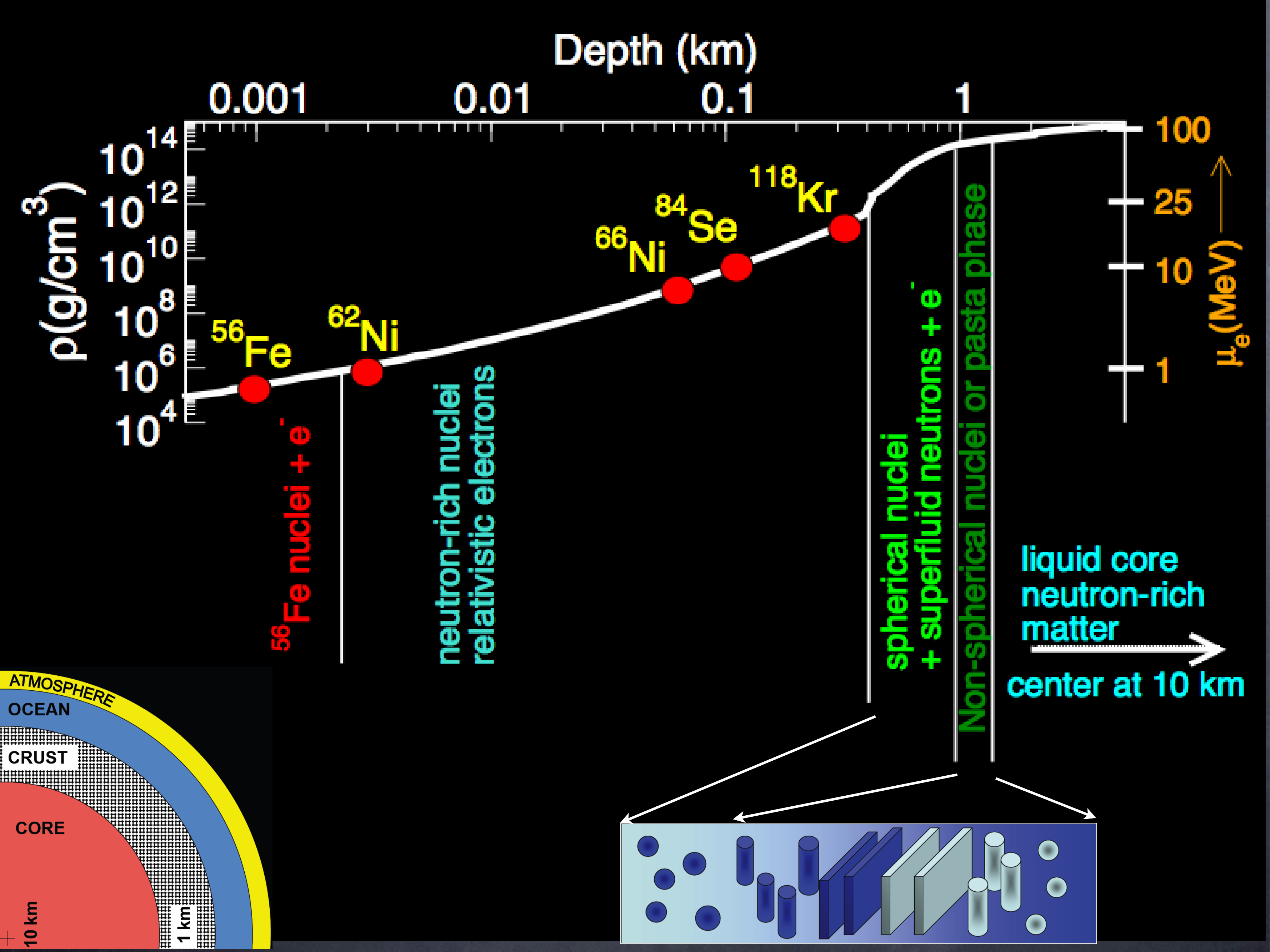} 
   \caption{Neutron star profile emphasizing the crust. The top x-axis shows the depth from the surface and left and right y-axes show the density and electron chemical potential, respectively.}
   \label{fig:Crust_Illustration}
\end{figure}

\lSect{composition}
The neutron star crust is composed of matter at sub-nuclear density ($ \rho \leq 10^{14}$ g/cm$^3$) and occupies the outermost 1-2 km region of the star.  In the outer crust, matter is composed of fully pressure ionized nuclei immersed in a degenerate electron gas. With increasing density, the electron Fermi momentum  $k_{\rm Fe}= (3 \pi^2 Z~n_I)^{1/3}$, where $Z$ and $n_I$ are the charge and number density of nuclei, increases rapidly and the reaction $e^-+p\rightarrow n+ \nu_e$ drives nuclei to become increasingly neutron rich. Eventually, at $\rho \simeq 4-7~\times10^{11}$ g/cm$^3$ the neutron fraction exceeds that which can be  bound in the nucleus and some neutrons drip out and become delocalized from nuclei. The fraction of unbound neutrons increases rapidly with density and they become superfluid due to strong attractive interactions. The illustration in Fig.~\ref{fig:Crust_Illustration} shows typical variation in composition, density and electron chemical potential in the neutron star as a function of depth from the surface. Here it is assumed temperature is small compared to the energy differences between the ground state nucleus and other excited or metastable states. Under these conditions a unique ground state nucleus exists at each depth and is referred to as cold catalyzed matter.  The region between neutron-drip, where the favored ground state nucleus is $^{118}$Kr, and the core of the neutron star is called the inner curst and was first studied in pioneering work by Baym, Bethe and Pethick \cite{Baym:1971pi} and by Negele \& Vautherin \cite{Negele:1973ve}. In the shallower regions of the inner crust, spherical neutron-rich nuclei are embedded in a neutron superfluid, but with increasing density non-spherical shapes (collectively referred to as "pasta" phases \cite{Pethick:1995}) are likely to appear (see inset in Fig.~\ref{fig:Crust_Illustration}).

The composition and structure of the inner crust  have been extensively studied using mean field models based on the Skyrme interaction or its relativistic generalizations. These studies have elucidated how the structure and composition are determined by two key driving forces: (i) the density dependence of the nuclear symmetry energy; and (ii)  the gradient or surface energy in heterogenous asymmetric matter. Although the suite of models have explored a range of predictions for the symmetry and gradient energy, these ingredients remain poorly constrained by both theory and experiment. Hence the spatial extent and composition of the denser regions where pasta phases are expected, are active areas of research and is discussed in several contributions in this book.
\begin{figure}[h]
   \centering
   \includegraphics[width=0.65\textwidth]{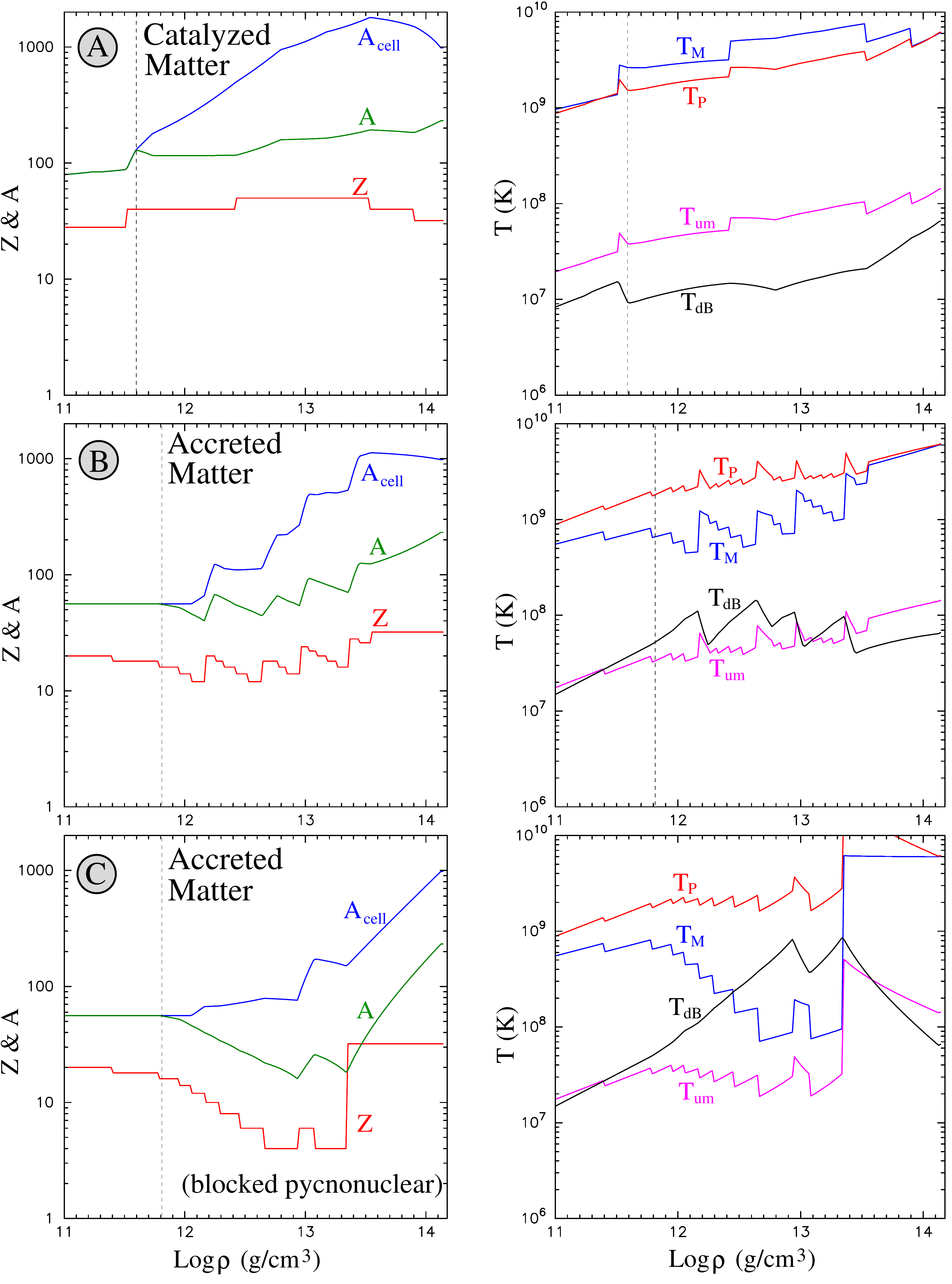} 
   \caption{Left Panels: Crust chemistry. The variation of $Z$, $A$  and $A_\mathrm{cell}$ with density. The
vertical dotted line show the neutron drip point $\rho_\mathrm{drip}$. Right panels show characteristic temperature scales discussed in \Sect{thermal}:   
   $T_\mathrm{M}$ is the classical ion melting temperature, for a critical $\Gamma \simeq 200$,
   $T_\mathrm{P}$ is the ion plasma temperature,
   $T_\mathrm{um}$ is the crystal umklapp temperature, and
   $T_\mathrm{dB}$ is the ion de Broglie temperature.
     }
   \label{fig:Chem}
\end{figure}

The composition in accreting neutron stars can be more complex since the crust here is formed almost entirely from accreted matter. Nuclear reactions that process this material have been studied in the literature and several relevant reaction pathways are now known [see \cite{Gupta:2008,Haensel:2008ly}],  but the rates of some crucial pycno-nuclear fusion reactions at the temperatures attained in the crust remain very uncertain\cite{Yakovlev:2006}. Thus, several metastable nuclei can coexist at the same depth in the crust depending on its formation history. In the outer crust, the diversity in nuclear species is expected to be quite large in systems where nuclear burning and rapid-proton (rp) capture reactions during an X-ray bursts can produce nuclei with large Z and A . With increasing depth a sequence of electron capture,  and neutron emission and absorption reactions tend to reduce both the diversity and the average $Z$ and $A$ of the nuclei encountered \cite{Gupta:2008}. These nuclei with lower Z and A are metastable because the fusion reactions needed to process them to the ground state are slow.   

The composition of matter is shown in the left panels of Fig.~\ref{fig:Chem} where $Z$ is the nuclear charge, $A=N+Z$ where $N$ is the number of neutrons in the nucleus with single particle energy $<0$, and $A_\mathrm{cell}=N+Z+N_{\rm out}$ is the total number of nucleons per unit cell (containing one nucleus) and $N_{\rm out}$ is the number of neutrons with  single particle energy $>0$ in the cell.  The panels show: (A) cold catalyzed matter where matter is in the ground state \cite{Negele:1973ve} ; (B) matter in accreting neutron stars where nuclear reactions including pycno-nuclear fusion process $^{56}$Fe at $\rho \simeq 10^9$ g cm$^{-3}$ to higher density  \cite{Haensel:2008ly}; and (C) matter in accreting neutron stars where pycno-nuclear fusion is assumed to not occur for $Z > 4$; are shown. In models (B) and (C) we use the composition of \cite{Negele:1973ve} for catalyzed matter for 
$\rho > 2\times 10^{13}$ g/cm$^3$.  In these plots we assume $Z$ changes abruptly while $A$ and $A_\mathrm{cell}$ vary smoothly, and the vertical dashed lines show the point at which neutron drip occurs.  


Unless stated otherwise, in this article we use natural units where the speed of light $c=1$, Planck's constant $\hbar=1$ and the Boltzmann constant $k_B=1$. Typically, we will quote energy in MeV and length in fermi and velocities in units of $c$. To convert to cgs units, quantities of interests such as the thermal conductivity and the specific heat per unit volume,  we note that $ {\rm MeV}^2 = 1.07\times10^{16} [\frac{{\rm ergs}}{{\rm cm}~{\rm s} ~{\rm K}}]$ and  ${\rm MeV}^3 = 1.81\times10^{16} [\frac{{\rm ergs}}{{\rm cm}^3~{\rm K}}]$.  
\section{Low energy excitations and thermal properties}
\lSect{thermal}
To describe the thermal properties we will dwell on a key simplification of dynamics at low temperature where only electrons and collective excitations or phonons are relevant.  As we shall see below the details of the complex nuclear ground state and the lattice structure are encoded in a few low energy constants which determine the propagation and interaction of phonons. Three of these modes are phonons of the lattice, and can be thought of a coherent motion of the clusters of protons that are located at the center of the neutron-rich nuclei. The forth mode is associated with neutron density oscillations and is called the superfluid phonon mode when neutrons are superfluid. The motion of proton clusters and neutrons are not independent, and a fraction of the neutron fluid is entrained in the motion of the proton clusters. The number of neutrons entrained by each proton cluster is in general different from $N$ and as we shall discuss below it is a key microscopic input for the low energy theory of phonons.  

\begin{figure}[htbp]
   \centering
   \includegraphics[width=0.8\textwidth]{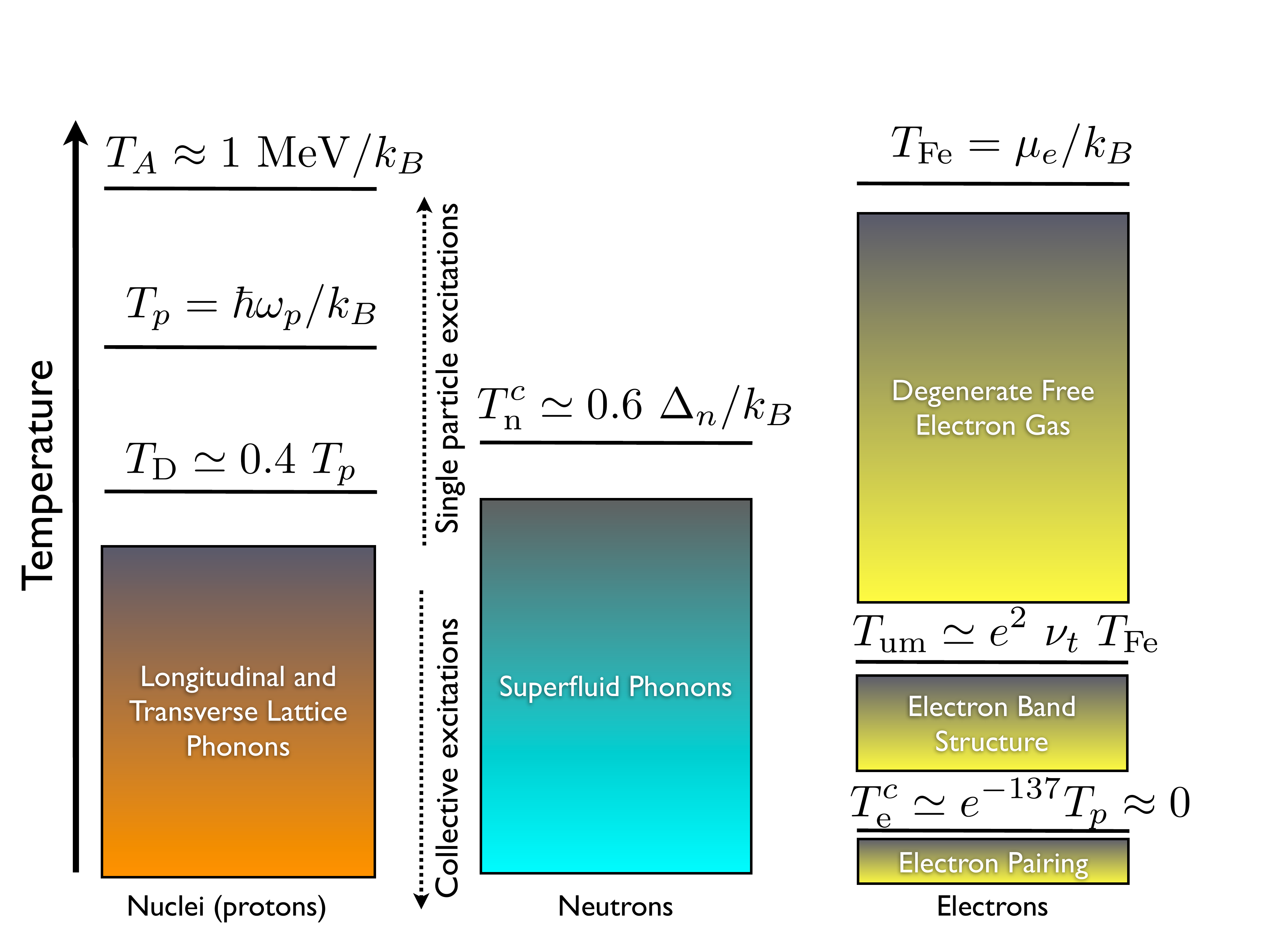}
   \caption{The regimes where collective excitations dominate over single particle excitation for ions and neutrons. Scales of relevance to electron dynamics are also shown.}
   \label{fig:Scales}
\end{figure}

To begin we shall briefly discuss the separation between the low and high energy scales for each of the three components in the inner crust shown in Fig.~\ref{fig:Scales}.
\subsection{Ions}
Since the electron screening momentum $k_{\rm TFe} = \sqrt{4 e^2/\pi}~ k_{\rm Fe}\lessapprox  (2a)^{-1} $  where $2a$ is inter-ion distance, ions crystallize at 
$T \lessapprox T_M \simeq Z^2 e^2/(\Gamma_\mathrm{M}~a)$, where $T_M$ is the melting temperature of the Coulomb solid. 
For a one component Coulomb system $\Gamma_\mathrm{M} =173$ but can be as large as 230 for a mixture a elements.
Written in terms of the chemical potential $\mu_e=k_{\rm Fe}$ for relativistic electrons,  
\be
T_M \simeq  3.4 \times 10^9 \left(\frac{Z}{40}\right)^{5/3}~\frac{\mu_e}{30 ~{\rm MeV}} ~~{\rm K} \,,
\ee 
which implies that the inner crust where $\mu_e\simeq 30-100$ MeV is a solid for  $T \le  10^{9}$ K.
 For ions, the intrinsic energy scale is the plasma frequency $\omega_p = (4\pi e^2 n_I~Z^2/M^*_I)^{1/2}$ where $n_I=3/4\pi~a^3$ is the number density of ions, and $Z$ and $ M^*_I $ are the charge and effective mass of the ions, respectively. In terms of $\mu_e$,
\be
\omega_p = T_p =
\sqrt{\frac{4 Ze^2~\mu_e^3}{3 \pi ~M^*_I}} \simeq 0.3~\sqrt{\frac{Z}{A^*}}~\left(\frac{\mu_e}{30~{\rm MeV}}\right)^{3/2}  {\rm MeV} \,, 
\ee
where $A^*\approx M^*/m_n$ is the effective number of nucleons that move (entrained) with each nucleus. 

Below the Debye temperature $T_{\rm D} \simeq 0.45 T_p$ one longitudinal and two  transverse phonons of the lattice capture all of the relevant ion dynamics. Uncorrelated single particle motion for ions does not emerge until  $T\gg T_p$. The variations of $T_\mathrm{M}$ and $T_p$ with density and composition are shown in Fig.~\ref{fig:Chem}. Generally ,we can expect that nuclear excitations are suppressed by shell gaps and pairing correlations. The evolution of these effects in the inner crust is yet to be understood. We denote this as $T_A\approx 1 $ MeV - a typical nuclear physics scale. It is possible that surface modes and low-lying collective excitations in the large nuclear structures encountered in the pasta-phases could have smaller energy.

\subsection{Electrons}
In terrestrial solids, the electron excitation spectrum is complex as band structure and Cooper pairing play a role at low temperature manifesting in insulating, metallic and superconducting properties. In contrast,  electronic structure in the crust is simpler because the electron Fermi momentum $k_{\rm Fe}$ is larger than all other relevant scales depicted in Fig.~\ref{fig:Scales}. 

Electrons are relativistic with $v_F\simeq1$ and the ratio $V_{\rm e-i}/\mu_e \simeq Z^{2/3}~e^2 \ll 1$ where  $V_{\rm e-i}=Ze^2/a$ is the interaction energy and $\mu_e= (9\pi~Z/4)^{1/3}/a$ is the  electron chemical potential. Consequently the Fermi surface is nearly spherical with minor distortions at intersections with Brillouin zone boundaries where coherent Bragg scattering leads to a band gap\cite{Kittel:1976}. These small patches on the Fermi surface become important only at low temperature when considering phonon mediated transitions between electrons.  As we discuss in 
\Sect{transport}, the gap affects the Umklapp process for $T \lessapprox T_{\rm um} =v_t~\delta_U$  
where $\delta_U$ is the band gap and $v_t$ is the velocity of transverse phonons \cite{RaikhYakovlev_82}. For nearly free electrons $\delta_U = V_{k_{Fe}}$, where $V_{k_{Fe}} \simeq 4 \pi Z e^2~n_I/k_{Fe}^2 =(4e^2/3\pi)~k_{Fe}$ is the Fourier component of the lattice potential at scale $k_{Fe}$ .  

Although electrons in the inner crust are as degenerate as terrestrial superconductors with $T/T_F \approx 10^{-5}-10^{-4}$, where $T_F=\mu_e$, the critical temperature is negligibly small because here electrons are relativistic. They move too quickly to adequately experience the attraction due to retardation effects in the electron-phonon potential, and  consequently the critical temperature $T^c_e \simeq  \omega_p~\exp{(-v_F/e^2)} \ll \omega_p$  is negligibly small \cite{Ginzburg:1969}.  Thus, the degenerate Fermi gas model provides an excellent description of electronic properties for $ T \le T_F$. In this regime, the density of states $N_e(0)= \mu_{e}^2/\pi^2$ is large and this greatly enhances their contribution to thermal and transport properties at low temperature.                       
\subsection{Neutrons}
\begin{figure}[t]
   \centering
   \includegraphics[width=0.4\textwidth]{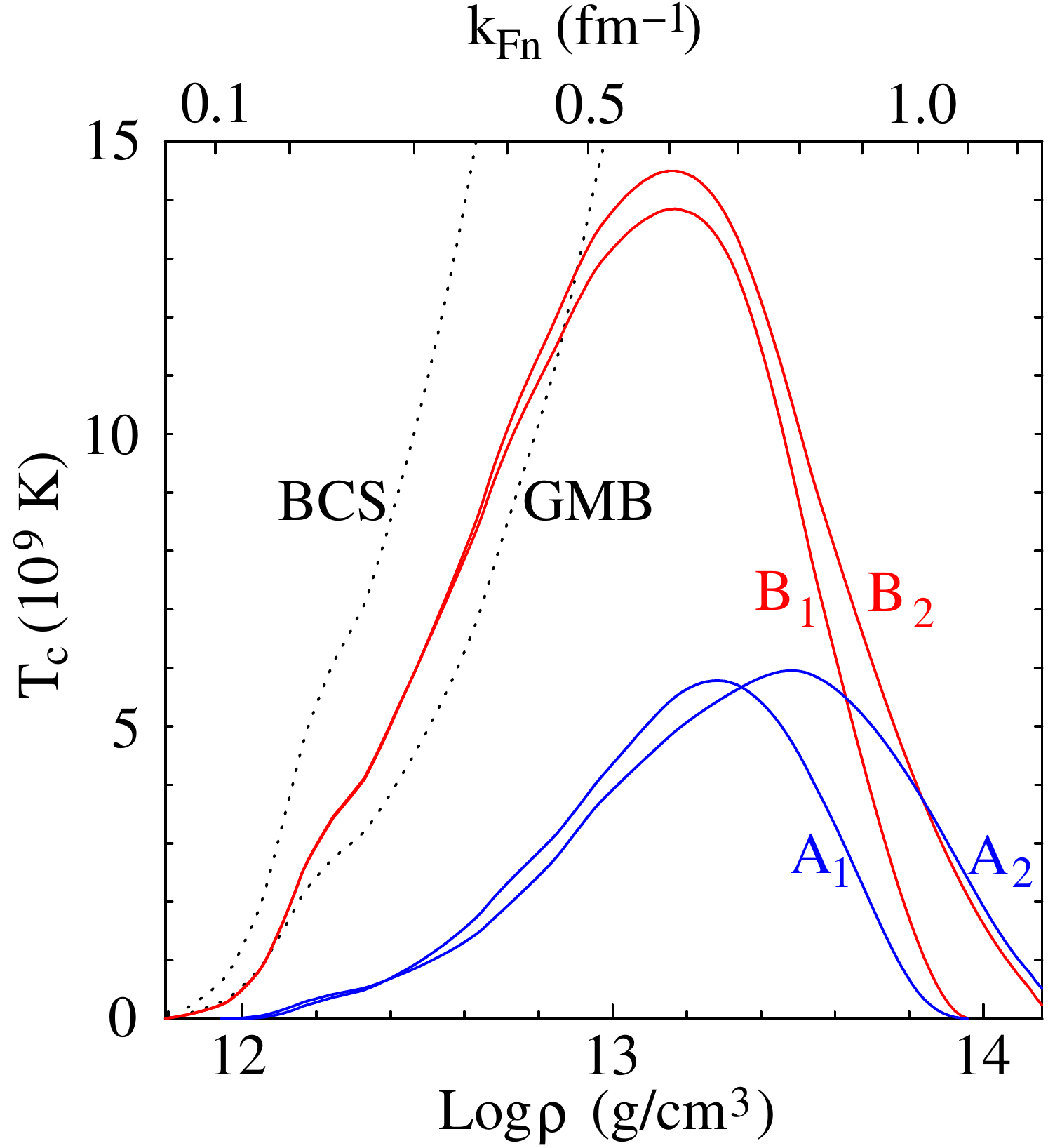}
   \caption{A sample of theoretical prediction for the neutron $^1$S$_0$ superfluidity critical temperature $T_c$.
   }
   \label{fig:Tc}
\end{figure}

Due to strong attractive interactions, neutrons in the inner crust form Cooper pairs and become superfluid. The gap in the single particle spectrum is denoted by $\Delta_n$ increases from zero at neutron drip to a maximum value $\approx 1$ MeV at a density $\rho \simeq 10^{13}$ g/cm$^3$ and decreases therafter.    The number of thermally excited neutron quasi-particles is exponentially suppressed when $T \ll \Delta_n $ and their contribution to thermal and transport properties is typically negligible. However, depending on the variation of the gap with density, a sizable fraction of the inner crust close to neutron drip and the vicinity of the crust-core interface can be normal in accreting neutron stars.  

In Fig.~\ref{fig:Tc} model predictions for the critical temperature $T_c = \Delta_n/1.76$ are shown where curves labelled "BCS" and "GMB" show the analytical results in the weak coupling valid in the limit $|a k_F| \ll 1$. In the Bardeen Cooper and Schrieffer (BCS) approximation $\Delta_\mathrm{BCS} = (8/e^2) \exp(\pi/2 a ~k_F) E_F$, with a scattering length $a = -18.5$, fm.  Corrections due to medium polarization which appear at the same order reduce the gap to $\Delta_\mathrm{GMB} = 1/(4e)^{1/3} \Delta_\mathrm{BCS}$ from \cite{Gorkov:1961kx}.
Curves labelled  "A1" and "A2" are examples of slowly growing $T_c$ at low $k_F$,  from \cite{Chen:1993ys} and \cite{Wambach:1993kl}, respectively.  Curves "B1" and "B2" mimic behavior predicted by strong coupling QMC calculations from \cite{Gezerlis:2008tg} and \cite{Gandolfi:2008hc} where the gap increases rapidly with density.  
In models labelled  "A1" and "B1" where gaps vanish at $\rho\simeq10^{14}$ g/cm$^3$.  For more details on the density and model dependence of the gap we refer the reader to  the chapter by Gezerlis and Carlson\cite{Gezerlis:2011uq} in this book. 
  
In the region where $T < T_c$ collective excitations of the neutron fluid called superfluid phonons, 
with a dispersion relation $\omega=v_\phi~q$, are 
the relevant low energy degrees of freedom. This mode corresponds to fluctuations of the phase of the superfluid condensate (and can be related to density fluctuations) and is the Goldstone mode associated with the spontaneous breaking of the global $U(1)$ symmetry in superfluid ground state (the Hamiltonian is invariant under arbitrary phase rotations of the fermion fields, but in the superfluid ground state is preserved only by discrete rotations of $\pi/2$). 

\subsection{Specific heat}
The electron contribution the specific heat (hereafter $C_v$ will represent the specific heat per unit volume) is given by 
\be 
C^e_v= \frac{1}{3}\mu_e^2~T\,, 
\label{eq:cve}
\ee
at low temperature. Band structure affects only negligible as only small regions of the Fermi surface are affected. At low-temperature when $ T \ll T_p$ electrons dominate, but as we discuss below the phonon contribution can become important in accreting neutron stars where $T \simeq 10^8-10^9$ K. For $T \lessapprox T_D$ the contribution from lattice phonons (lph) is given by 
\be
C^{\rm lph}_v = \frac{2\pi^2}{15}~\left(\frac{T^3}{v_l^3} + \frac{2~T^3}{v_t^3}\right) \,,
\label{eq:cvlph}
\ee  
and $v_l$ and $v_t$ are velocities of the longitudinal, and transverse lattice phonons, respectively. In a model where the strong interaction between the neutron superfluid and the ion lattice is ignored it is simple to calculate these velocities. The speed of longitudinal lattice vibrations is approximated as $v_l = \sqrt{K_{\rm ion-e}/\rho}$  where $K_{{\rm ion-e}} = \rho(\partial(P_{\rm ion} +P_e)/\partial \rho)$ is the bulk-modulus of the electron-ion system  and  the ion mass density $\rho  = A m_n~n_I$  where A is the number of bound nucleons in the ion.  Since $P_e \gg P_{\rm ion} $, we can write 
\be
v_l=\sqrt{\frac{\partial P_e} {\partial\rho}}=\frac{\omega_p}{k_{\rm TFe}} \,,
\ee
which is usually referred to as the Bohm-Staver sound speed. The velocity of the transverse lattice mode is related to $\mu$, the shear modulus of the lattice,  and is given by  
\be 
v_t =    \sqrt{\frac{\mu}{\rho}} = \alpha \frac{\omega_{p}}{q_{\rm D}}\,,
\ee 
where $q_{\rm D}=(6 \pi^2 n_I)^{1/3}$ is the ion Debye momentum, and the constant $\alpha \simeq   0.4$ is obtained by numerical calculations of Coulomb crystals  \cite{1992Natur.360...48C}. 
Further, since 
\be 
\frac{q_{\rm D}}{k_{\rm T	Fe}} = \sqrt{\frac{\pi}{4e^2}}~\left(\frac{2}{Z}\right)^{1/3}  \gg 1
\ee
we have $v_l  \gg  v_t$ and the contribution from longitudinal modes  to $C_v$ in Eq.~\ref{eq:cvlph} in negligible. Thus the lattice contribution can be written in the familiar form  
\be
C^{\rm lph}_v= n_i~\frac{12\pi^4}{5}~\left( \frac{T}{T_{\rm D}}\right)^3 \,,
\ee
 where $T_{\rm D}=(3/2)^{1/3}v_t~q_{\rm D} \simeq 0.45 T_p $ is the Debye temperature of the ion lattice.  This low temperature form of the specific heat provides an excellent approximation in Coulomb solids up to $T \le T_p/50$ but fails when $T \ge  T_p/10$ \cite{Baiko:2001qf}.  

To calculate the neutron contribution to $C_v$ we first note that there are two distinct regimes. In the normal phase when $T \ge T_c$ the neutron contribution is large and is given by 
\be
C^{\rm neutron}_v = \frac{1}{3}~m_n~k_{\rm Fn}~T \qquad \qquad \quad (T>T_c)
\label{eq:cv_neutron} 
\ee
This normal contribution can become important in the vicinity of neutron drip where $T>T_c$, and at the crust-core boundary. In the superfluid phase when $T \ll T_c$ the neutron single particle excitations are strongly suppressed and
\be
C^{\rm neutron}_v \approx  \frac{1}{3}~m_n~k_{\rm Fn}~T~\exp{\left(\frac{-\Delta_n}{T}\right)} \quad (T\ll T_c) \,.
\label{eq:cv_neutron2} 
\ee 
which is usually negligible. The four models for the gap in Fig.~\ref{fig:Tc} allow us to explore the effect of pairing on the neutron specific heat. In models A1 and A2 we have a thick shell of normal neutrons above the drip point, while models A1 and B1 predict a thick layer of normal neutrons at the highest densities. 
Modifications to this simple picture of pairing in uniform neutron matter due to the presence of the nuclei are discussed in this book 
in the chapter by N. Sandulescu \& J. Margueron \cite{Margueron:2012ph}. Further, we briefly note that like in the case of electrons, coherent Bragg scattering of neutrons by the lattice lead to band structure effects that modify the shape of 
the Fermi surface, still Eq.~\ref{eq:cv_neutron} is an excellent approximation to $C_v$ in normal phase for reasons described in \cite{Chamel:2008ju}.

Elsewhere in the crust where $T\ll T_c$ the relevant neutron contribution is from superfluid phonons, i.e., collective instead of single particle excitations, and is given by 
 \be 
 C^{\rm sph}_v = \frac{2\pi^2}{15}\frac{T^3}{v_\phi^3} 
\label{eq:cvsph} 
\ee
where 
\begin{equation} 
v_{\phi} = \sqrt{\frac{n_f}{m_n~f^2_\phi}}
\;\;\;\;
\left(\mathrm{with} \;f^2_\phi = \frac{\partial n_f}{\partial \mu_n} \, ,
\;\;\; \mathrm{see\;  \Sect{entrain}}
\right) \,
 \label{eq:vsuper}
\end{equation}
is the superfluid phonon velocity, $n_f$, $\mu_n$ and $m_n$ are the number density, chemical potential and mass of the free neutrons, respectively. For  weakly coupled systems $v_{\phi} = v_{\rm F}/\sqrt{3}$ where $v_{\rm F}$ is the Fermi velocity. In most of the inner crust $ v_\phi \gg v_t$ (see Fig.~\ref{fig:velocity}) and hence their contribution 
to  the heat capacity is negligible except perhaps in a sliver where $v_\phi \simeq v_t$ and $T \le T_c$.   
\begin{figure}[h]
   \centering
   \includegraphics[width=0.95\textwidth]{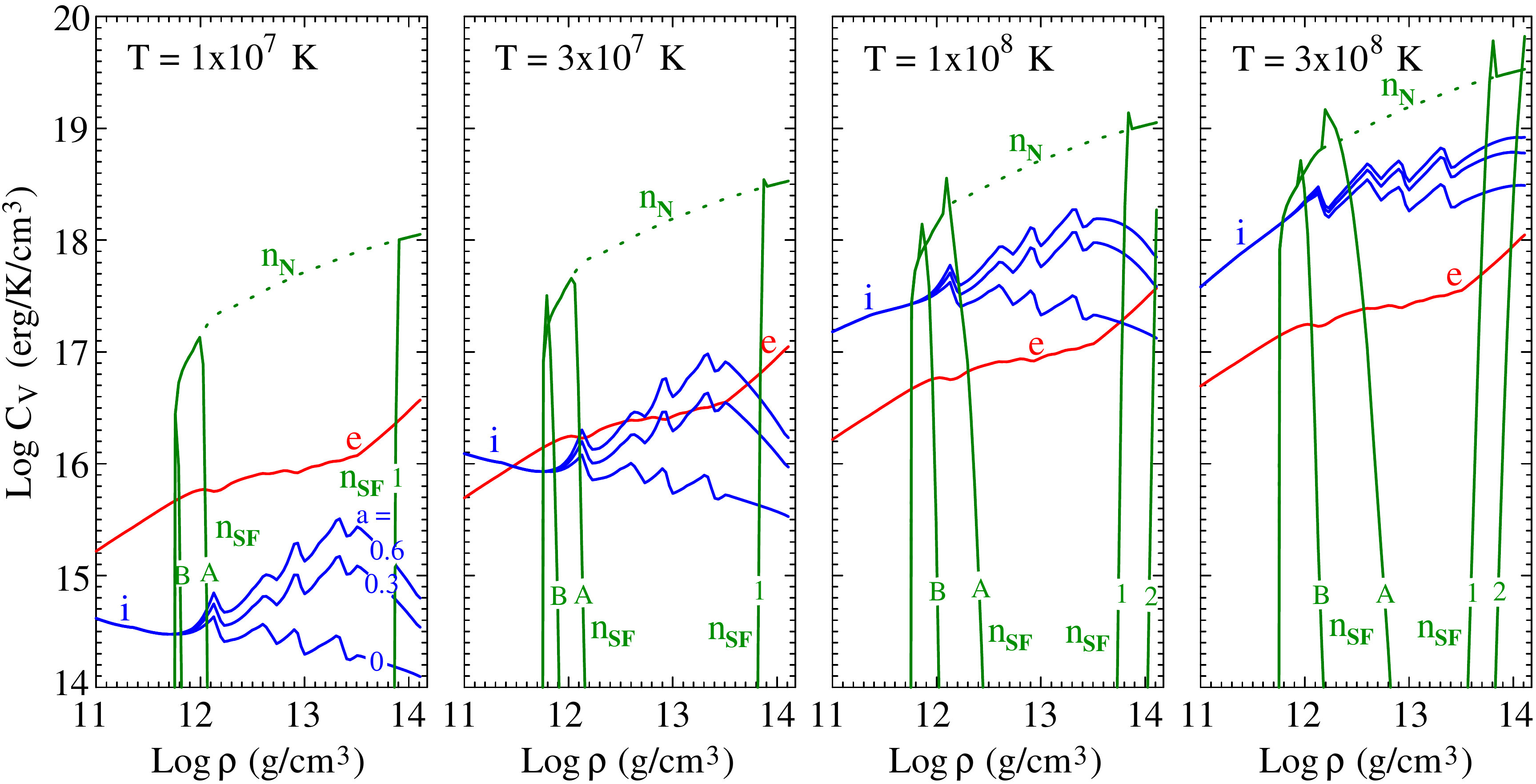} 
   \caption{Specific heat of ions, 
   electrons, and for neutrons with (labelled $n_\mathrm{SF}$) and without the effects of the superfluid gap (labelled $n_\mathrm{N}$) are shown for four representative temperatures. }
   \label{fig:Cv}
\end{figure}

The specific heat due to these components is shown in Fig.~\ref{fig:Cv}. The ion contribution for $T \lessapprox 0.1 T_p$ varies as $T^3$ and is to very good approximation given by $C^{\rm lph}_V$, while  electron contribution is linear in T and dominates at low temperature. As mentioned earlier, the neutron contribution is sensitive to the variation of the  $^1S_0$ gap. 
To illustrate this we show predictions representative of two distinct possibilities corresponding to the behavior akin to models
 A's and B's of Fig. \ref{fig:Tc}. In models labelled A1 and A2 the neutron contribution is relevant over an extended region near neutron drip, while in models A1 and B1 the contribution is relevant in the region close to the crust-core boundary.
The contribution of the superfluid mode is negligible and is not shown. The ion contribution in the phonon regime $T<T_D$ changes quite significantly with the effective mass of the nucleus
since $C^{\rm lph}_V \propto M_I^{*\, 3/2}$: its increase from  $a=0$ to $30\%$ and $60\%$, where $a$ is the fraction of unbound neutrons entrained by each nucleus, is illustrated in Fig. \ref{fig:Cv}. We now turn to briefly describe the low energy theory where effects due to entrainment are taken into account consistently.  
\subsection{Entrainment and mixing}
\lSect{entrain}
In the preceding discussion the interaction between the neutron superfluid and the lattice was ignored. Because neutrons and protons interact strongly this is difficult to justify and recent work in \cite{Cirigliano:2011} describes a first attempt at including these interactions. Using the techniques of low energy effective field theory, a theory of lattice and superfluid phonons was formulated where the relevant fields $\phi$ and $\xi^a$ correspond to the phase of the neutron field and the displacement vector of the proton clusters from their equilibrium positions, respectively. The symmetries associated with translation and number conservation require that the low energy theory be invariant under the transformation $\xi^{a=1..3}({\bf{r}}, t)\rightarrow \xi^{a=1..3}({\bf{r}}, t) + a^{a=1..3} $ and $\phi({\bf{r}}, t)\rightarrow \phi({\bf{r}}, t) + \theta$ where $a^{a=1..3} $ and $\theta$ are constant shifts. Since the theory must respect these symmetries, it implies that the Lagrangian can contain only spatial and temporal gradients of these fields. This gradient expansion enables us to organize the calculation of any low energy process in a well controlled expansion in powers $p/\Lambda$ where p is the momentum (involved in the process) and $\Lambda$ is an intrinsic high-momentum scale associated with the ground state.  Assuming that the ground state must posses a cubic symmetry, the 
quadratic part of the effective Lagrangian density 
\begin{eqnarray}
{\cal{L}_I} &=& \frac{f_\phi^2}{2}(\partial_0\phi)^2 - \frac{v_\phi^2f_\phi^2}{2}(\partial_i\phi)^2
+\frac{\rho_I}{2}\partial_0\xi^a\partial_0\xi^a - \frac{1}{4} {{\mu_I}}(\xi^{ab}\xi^{ab}) -
\frac{{K_I}}{2}(\partial_a\xi^a)(\partial_b\xi^b) \nonumber \\
&-& \frac{\alpha}{2}\sum_{a=1..3}(\partial_a\xi^a\partial_a\xi^a) +{g_{\rm
mix}f_\phi\sqrt{\rho_I}}~\partial_0\phi\partial_a\xi^a + \cdots~\label{eq:phenomLagrang}\;, 
\end{eqnarray}
where higher order terms involve higher powers of the gradients of these fields, and 
$\xi^{ab} = 
(\partial_a\xi^b+\partial_b\xi^a) - \frac{2}{3}\partial_c\xi^c\delta^{ab}$. 

If the superfluid and the solid are uncoupled, 
the low energy coefficients (LECs) appearing above, such as $\rho_I, {\mu_I},{K_I}$ are the mass density, the shear modulus, and the compressibility of the solid, respectively, and $f_\phi$ and $v_\phi$ of the superfluid are as defined earlier in Eq.~\ref{eq:vsuper}. They
determine the velocities
\begin{equation} 
 v_l = \sqrt{\frac{{K_I}+(4/3){\mu_I}}{\rho_I}}\,,\quad v_t = \sqrt{\frac{{\mu_I}}{\rho_I}}\,,\quad {\rm and}\quad v_{\phi} = \sqrt{\frac{n_f}{m_n~f^2_\phi}}\,,
 \label{eq:velocities}
\end{equation}
respectively, where $n_f$ now corresponds to the number density of free neutrons and $\rho_I=A m_n~n_I$ the mass density of the nuclei.  In the presence of strong coupling between the solid and superfluid these coefficients are modified.  As the proton clusters move, they drag along neutrons from the fluid and this will modify the mass density involved in lattice fluctuations.  Correspondingly, the coefficient $\rho_I$ appearing in Eq.~\ref{eq:phenomLagrang} and the lattice velocities will differs from the usual mass density of the pure lattice component due interactions that entrain the superfluid  \cite{Carter:2004pp,Carter:2006}. Naively, one may associate the number of neutrons entrained by each proton cluster to correspond to the number of bound neutrons  with single-particle energy less than zero. However, this typically underestimates the mass-density associated with lattice motion. Calculations indicate that a large fraction of neutrons with positive single particle energies are entrained due to coherent  Bragg scattering of neutrons from the lattice \cite{Chamel:2005}.  Denoting this number density of entrained neutrons as $n_b$ we can write  $\rho_I=(n_b+n_p)m_n$ where $n_p$ is the average proton number density in the unit cell.   

It is convenient to express $n_b$ in terms of the neutron effective mass $m^*$ calculated from band structure studies outlined in \cite{Chamel:2005} 
\be 
\frac{n_b}{n_I} = A_{\rm cell } \left( 1-\frac{m_n}{m^*} \right)~+~A~\frac{m_n}{m^*}-Z=(A^*-Z)\,, 
\label{eq:nb} 
\ee 
where $A_{\rm cell}\simeq \rho/m_n$ is the total number of nucleons in the unit cell, $A$ is the number of nucleons bound in the nucleus and $A^*$ is the effective number of nucleons that move with the nucleus:
\be
A^* = A + a(A_\mathrm{cell} - A) \;\;\;\; \mathrm{with} \;\;\;\; a= 1-\frac{m_n}{m_n^*}\,. 
\ee
Representative values of $A_\mathrm{cell}$, $A$, $Z$ in the inner crust as function of  total mass density $\rho$ are shown Fig.~\ref{fig:Chem} .  Calculations reported in \cite{Chamel:2005} indicate that $m^*$ is typically in the range of $(3-6)~m_n$ but can be as large as $15~m_n$ is some regions where coherent Bragg scattering is most efficient. From Eq.~\ref{eq:nb} we see that when $m^* \gg m_n$ a large fraction of neutrons are entrained by nuclei.

Current conservation implies that the number of neutrons that move freely as part of the superfluid is correspondingly reduced and this is denoted by $n_f=(n_n-n_b)$ where $n_n$ is the total neutron density \cite{Pethick:2010}. This repartition of neutrons between the lattice and the superfluid will modify the LECs appearing in Eq.~\ref{eq:phenomLagrang}. Now $\rho_I=(n_b+n_p)m_n$ and $n_f=(n_n-n_b)$ in Eq.~\ref{eq:velocities}. The thermodynamic derivates needed to define the lattice compressibility $K$ the shear modulus $\mu$ also contain the effects of the neutron-proton interactions as they are accounted for in the calculation of the ground state properties. The specific thermodynamic derivatives that define the LECs of the mixed system are explicitly given in \cite{Cirigliano:2011}. 

\begin{figure}[htbp]
   \centering
   \includegraphics[width=0.60\textwidth]{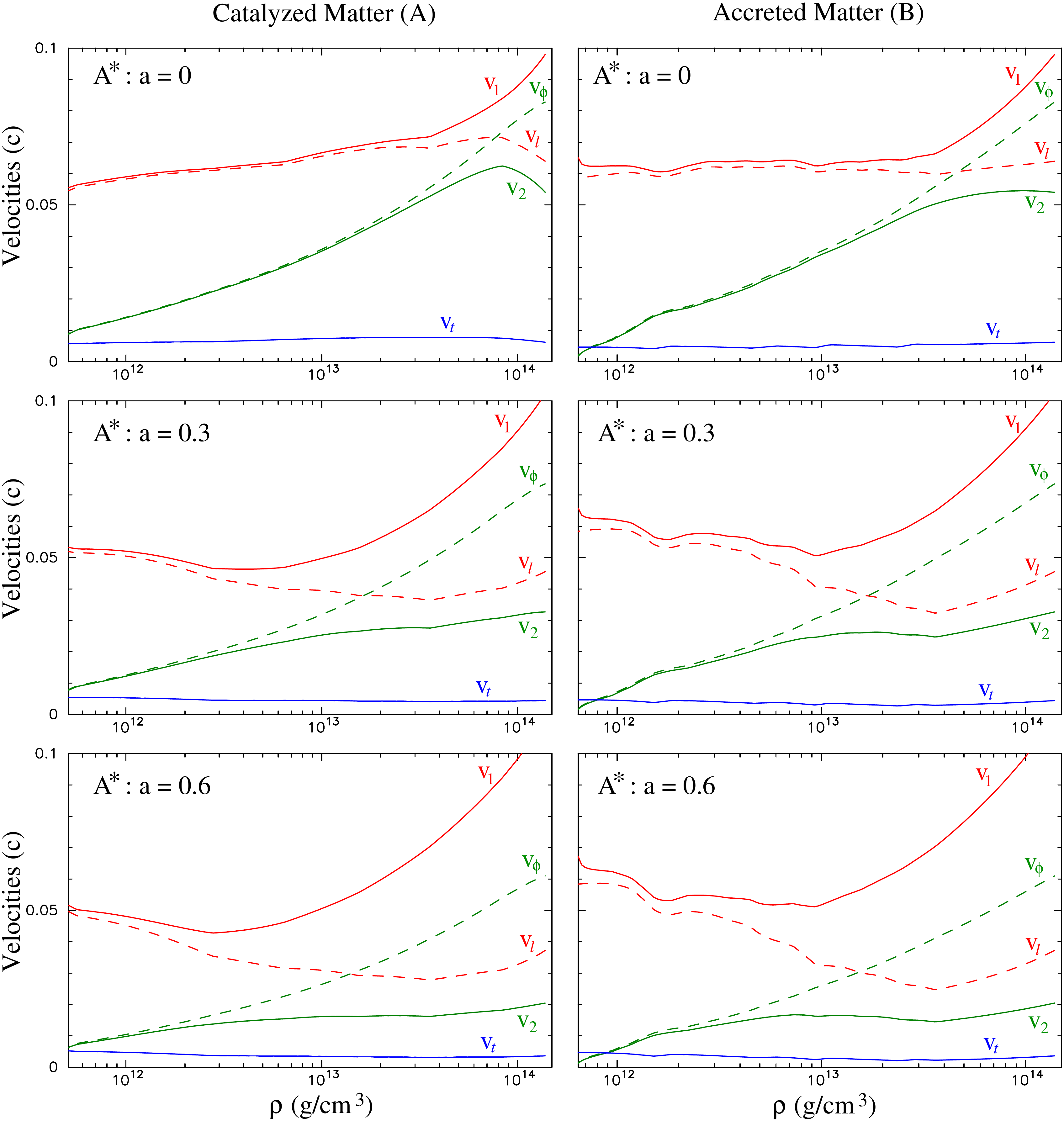}
   \caption{Velocities of phonons in the inner crust for two chemical compositions from Fig \ref{fig:Chem} and
   three values of the nuclei effective mass $A^*$.
   }
   \label{fig:velocity}
\end{figure}

Interactions between neutrons and protons leads to mixing between the modes of the lattice and the superfluid. The extent of this mixing is controlled by the dimensionless coefficient 
\begin{equation}
g_{\rm mix} = \frac{1}{f_\phi~\sqrt{\rho_I}} ~\left( n_b - n_p \frac{\partial n_n}{\partial n_p} \right) \,, 
\label{eq:gmix}
\end{equation}  
where the first term in the parenthesis arises due to the interaction between neutron and proton velocities (entrainment) and second term is due to the interaction between neutron and proton densities \cite{Cirigliano:2011}. Estimates indicate  $\partial n_n / \partial n_p <  n_b/n_p$ and mixing is mostly determined by the term proportional to $n_b$  \cite{Pethick:2010,Cirigliano:2011} and approximately we can write   
\be 
g_{\rm mix} \simeq \frac{n_b~v_{\phi}}{\sqrt{n_f (n_b+n_p)}} \,. 
\ee
Mixing implies that the longitudinal eigenmodes are superpositions of the longitudinal lattice and superfluid phonons. The velocity of these eigenmodes is  given by
\begin{equation}
v_{1,2} = \sqrt{\frac{X}{2}\left(1\pm \sqrt{1-\frac{4 v_l^2 v_\phi^2}{X^2}}\right)}
\end{equation} 
where $X=g_{\rm mix}^2+v_l^2 + v_\phi^2$ and $v_l$ and $v_\phi$ are defined in Eq.~\ref{eq:velocities}. The velocity of the eigenmodes for the crustal compositions of catalyzed and accreted matter shown in panels (A) and (B) of Fig.~\ref{fig:Chem} are plotted in Fig.~\ref{fig:velocity}. The dashed curves show results for $v_l$ and $v_\phi$ without mixing and they cross at $\rho \simeq 10^{13}$ g/cm$^3$. In this resonance region mixing is large and level repulsion can be significant. Away from resonance, the eigenmodes contain only small admixtures: below  $\rho \simeq 10^{13}$ g/cm$^3$ the mode labelled $v_2$ is predominantly the superfluid mode and above it is predominantly the lattice mode. In these calculations  we have neglected the second contribution in Eq.~\ref{eq:gmix} to $g_{\rm mix}$ and the value of $n_b$ was chosen somewhat arbitrarily to reflect the range of $m^*$ predicted in \cite{Chamel:2005}. The panels show results for three values of $ g_{\rm mix}$ chosen to reflect different fractions $a=0,~ 30\%$ and $60\%$ of unbound neutrons in the cell entrained by each nucleus. Transverse modes are unaffected by mixing at leading order but are affected by entrainment. Its variation in the crust for different values of $a$ is also shown in Fig.~\ref{fig:velocity}. Despite strong mixing $ v_t \ll v_{1} ~{\rm or}~v_2$, and transverse modes will continue to be dominate the  specific heat.

\section{Transport Properties} 
\begin{figure}     
\centering
   \includegraphics[width=0.5\textwidth]{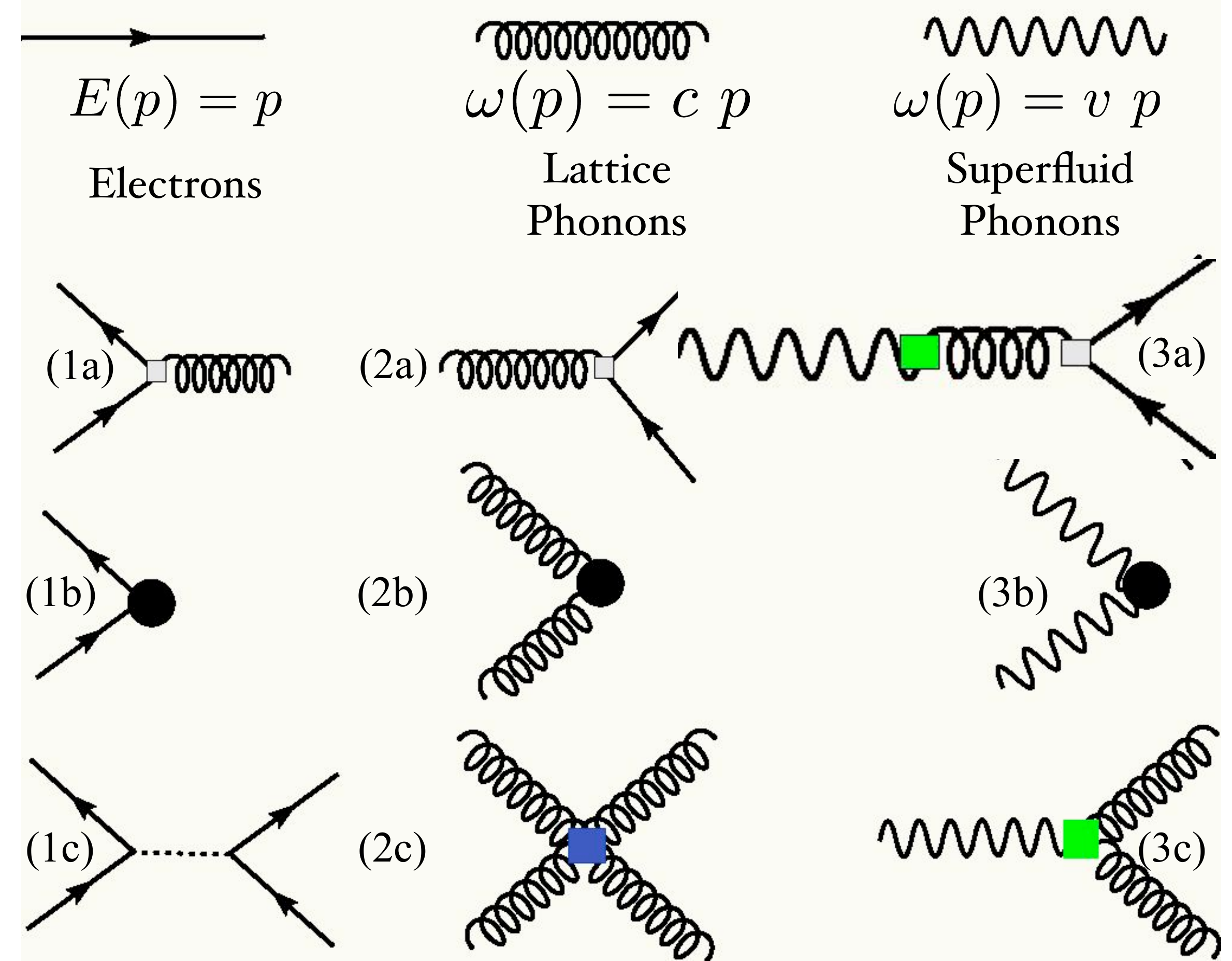} 
   \caption{Feynman diagrams indicating the various scattering and dissipative processes involving electrons, lattice phonons and superfluid phonons.}
   \label{fig:Feyn_Diagrams}
\end{figure}

\lSect{transport}
The electron and phonon thermal conductivity can be written as $\kappa =  C_v~v~\lambda/3$ where $C_v$ is their specific heat, $v$ is their velocity, and $\lambda$ is the transport mean free path. Using Eqs.~\ref{eq:cve} \& \ref{eq:cvlph} the electron and phonon conductivities are  
\begin{equation}
\kappa_e = \frac{1}{9}\mu_e^2~T~\lambda_e\,, \qquad \kappa_{\rm ph_i} =\frac{2\pi^2}{45 ~v_i^2}~T^3~\lambda_{\rm ph_i} 
\label{eq:kappa}
\end{equation}  
 where electrons are relativistic ($v=1$) with mean-free path $\lambda_e$, and the phonon contribution is for each phonon type with velocity $v_i$ and mean free path $\lambda_{\rm ph_i}$. Since $\mu_e \gg T$, electrons dominate at low temperature but phonon contributions can become relevant at high temperature when $\lambda_{\rm ph_i} \gtrapprox  (\mu_e/T)^2~v_i^2~\lambda_e$ or when the magnetic field is large enough to restrict electron motion \cite{Chugunov:2007,Aguilera:2008ed}. Phonon velocity was discussed in \Sect{thermal}, we now turn to discuss scattering and absorption processes that determine their mean free path. Feynman diagrams for relevant interactions are illustrated in Fig.~\ref{fig:Feyn_Diagrams} and in the following we briefly discuss the most important of these processes in the inner crust. 
 
\subsection{Electron-phonon processes} 
\lSect{elec_phon}
In its general form, the electron mean free path relevant for the thermal conductivity due to electron-ion scattering is given by 
\be
\lambda_e^{-1} = \frac{Z^2e^4}{4\pi \mu_e^2} \int_0^{2k_{Fe}} dk ~k^3~|\tilde{V}(k)|^2 \int_{-\infty}^{\infty} d\omega~{\cal F}(\beta \omega) S(\omega,k) ~g_\kappa(k,\beta \omega)
\label{eqn:Lambda_e}
\ee
where
\be
 g_\kappa(\beta\omega,k) = 1+\left(\frac{\beta \omega}{\pi}\right)^2 \left(3 \frac{k_{Fe}^2}{k^2}- \frac{1}{2}\right) \,,
 \;\;\;\; \;\;\;\;  
{\cal F}(\beta \omega) = \frac{\beta \omega}{\exp{(\beta \omega)}-1}
\label{eqn:gkappa}
\ee
and the dynamical structure factor $S(\omega,k)$ embodies all relevant dynamics of the strongly coupled system of ions \cite{FlowersItoh:1976}. 
Here, $\omega,k$ are the energy and momentum transfer.
The function $\tilde{V}(k) =F_Z(k)/(k^2+k_{\rm TFe}^2)$ characterizes the screened electron-ion interaction in momentum space  where $k_{\rm TFe}^2=4 e^2 k_{Fe}^2/\pi$ and $F_Z(k)$ is the charge form factor of the nucleus. 

Pauli blocking restricts $  \omega \simeq T \ll \mu_e$, and when $S(\omega,k)$ contains most of its strength in the region $\omega \lessapprox  3T$ the conductivity can be expressed in terms of the static structure function $S(k)= \int d\omega~S(k,\omega)$. However, $S(\omega,k)$ has strength at $\omega \simeq \omega_p$ and $\lambda_e$ cannot be calculated in terms of $S(k)$ when $T < T_p$.  Here, the frequency dependence of the dynamic structure factor is needed but this is generally difficult to calculate in strongly coupled quantum systems. Fortunately, when $T<T_D$ phonons are the only relevant degrees of freedom and electron scattering is dominated by the emission or absorption of phonons \cite{Ziman:1960}. In this case, $S(\omega,k)$ is simpler and is characterized by discrete peaks at $\omega = v k$ associated with the excitation of phonons with velocity $v$. 

\begin{figure}[t]
   \centering
   \includegraphics[width=0.4\textwidth]{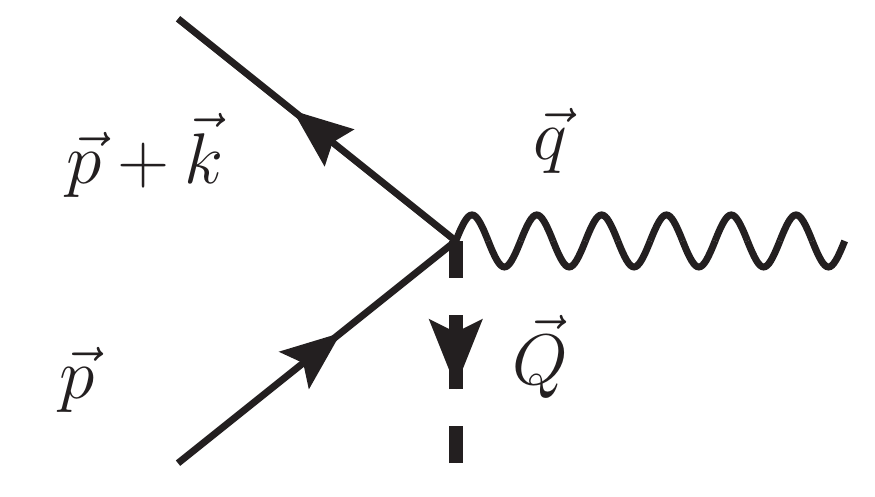} 
   \caption{Feynman diagram for the Umklapp process. The double dashed line represents recoil-free momentum transfer $\vec{Q}=\vec{k}-\vec{q}$ to the lattice, and $|\vec{q}|<q_D$ lies in the first  Brillouin zone.}
   \label{fig:Umklapp}
\end{figure}

In the low-energy theory, the interaction between electron and phonons is described by the Lagrangian density  
\begin{equation}  
{\cal L}_{\rm e-ph} = \frac{1}{f_{\rm eph}} \psi_e^\dagger \psi_e \partial_i \xi_i\quad {\rm where} \quad  {f_{\rm eph}}= \frac{\sqrt{\rho}~k_{\rm D}^2}{4 \pi Z e^2~n_I}
\label{eq:Lag_ep}
\end{equation} 
is related to electron-phonon coupling constant \cite{FetterWalecka:2003}, $ \psi_e$ is the electron field and $\xi_i$ is the ion displacement (phonon) field discussed in \Sect{thermal}. This form of the interaction applies to normal processes, where the momentum transfer $k < q_D $ and displacements correspond to excitation of longitudinal phonons. However, since $k_{Fe}/q_D=(Z/2)^{1/3} > 1$ large angle electron scattering with $k >  q_D $ is possible. This Umklapp process is depicted in Fig.~\ref{fig:Umklapp} where the electron simultaneously Bragg scatter off the lattice and excite a phonon. Elastic Bragg scattering (without phonon emission) however does not contrbute because electrons are eigenstates of the lattice potential. Further, unlike normal processes where only longitudinal modes are involved, Umklapp scattering is dominated by the emission or absorption of transverse phonons \cite{FlowersItoh:1976,YakovlevUrpin:1980}. 

The dynamic structure factor for  single-phonon emission and absorption including Umklapp shown in Fig.~\ref{fig:Umklapp} is given by 
\begin{equation}  
S(\omega,k)   = \frac{n_I}{M_I}\sum_i \sum_Q \frac{(\vec{k}.\hat{\epsilon_i})^2}{2~\omega}\left(\frac{\delta(\omega-v_i~q)}{1-\exp{(-\beta \omega)}} +\frac{\delta(\omega+v_i~q)}{\exp{(-\beta \omega)}-1}\right)  \delta^3(\vec{k}-\vec{Q}-\vec{q})\,,
\label{eq:swq_1ph}
\end{equation} 
where the first and second terms in parenthesis represent phonon emission and absorption, respectively \cite{FlowersItoh:1976} . The phonon momentum is restricted to the first Brillouin zone $q<q_D$, and sums are over all reciprocal lattice vectors or lattice momenta $\vec{Q}$, and the longitudinal and transverse phonon states with polarization vector $ \hat{\epsilon_i}$ and velocity $v_i$.  Using Eq.~\ref{eq:swq_1ph} and the delta functions to perform the integration over ${\bf k}$ and $\omega$, the electron mean free paths in Eq.~\ref{eqn:Lambda_e} can be written as 
\begin{eqnarray} 
\frac{1}{\lambda^{\rm ph}_e}&=&\frac{2\pi^2e^2~\omega_p^2}{\mu_e~T}~\sum_i~{\cal K}^{\rm(i)}(T,v_i) \,, \quad {\rm where} \,,\\
\label{eq:ltv}
{\cal K}^{\rm(i)}(T,v_i)&=&\sum^{{\cal P}<2k_{Fe}}_Q~\int_0^{q_{\rm D}}\frac{d^3q}{(2\pi)^3}~\tilde{V}({\cal P})\frac{{\cal P}(\vec{\cal P}.\hat{\epsilon_i})^2(1-{\cal P}^2/4k_{Fe}^2)~g_\kappa(\beta v_i q,\cal P)}{(\exp{(\beta v_i q)}-1)(1-\exp{-(\beta v_i q)})}\,,
\label{eq:Ktv}
\end{eqnarray}
and $\vec{\cal P}=\vec{q}+\vec{Q}$.  To unravel the dependence on the temperature and the phonon velocity we examine two limitings forms of the function ${\cal K}^{\rm(i)}(T,v_i)$. First, when $2k_{Fe} \gg q_{\rm D}$, the dominant contribution comes from the Umklapp and we can set $\vec{\cal P}= \vec{Q}$ in evaluating ${\cal K}^{\rm(i)}(T,v_i)$. In this case, from the RHS of Eq.~\ref{eq:Ktv} it is easy to deduce that 
\begin{equation}
\lim_{Q \gg q} {\cal K}^{\rm(i)}(T,v_i) \propto \frac{T^3}{v_i^3}\,. 
\end {equation} 
In the opposite limit, when only the normal process involving longitudinal lattice modes contribute we can set $ \vec{Q}=0$ in the RHS of Eq.~\ref{eq:Ktv} to find that 
\begin{equation}
\lim_{Q=0}{\cal K}^{\rm(i)}(T,v_l) \propto \frac{T^4}{v_l^4}\,. 
\end {equation} 

At very low temperature, the band gap in the electron spectrum suppress Umklapp processes. As mentioned in \Sect{thermal}, coherent Bragg scattering by the lattice will distort the electron Fermi surface for momenta that can coincide with the reciprocal lattice vectors $Q$. Here, the spectrum will differ due to a band gap $\delta_U \simeq  (4 e^3/3\pi)~ k_{\rm Fe}$.  Although distorted patches on the Fermi surface occupy only a small fraction of the total area, these regions are important for Umklapp transitions. To understand this suppression consider the case when the phonon momentum $q \approx 0$. In this limit, large angle electron Umklapp scattering with $\vec{k} \simeq \vec{Q}$ can only involve electrons on these patches. However, at low temperature the gap will suppress such transitions unless the phonon momentum $q \ge \delta k $ where  $\delta k\simeq \delta_U/v_{\rm Fe}$ can "steer" electrons away from these patches. For transverse thermal phonons $q \simeq  3 T/v_t$  and the condition on the phonon momentum implies that Umklapp occurs for $ T \ge T_{\rm um}$ where $T_{\rm um} =   (4 e^3/9\pi)~v_t~ k_{\rm Fe}$.   

From the preceding discussions we can conclude that for $T>T_{\rm um}$ the mean free path ${\lambda^{\rm ph}_e} \propto v_t^3/T^2$ since $v_ t \ll v_l$. For  $T \ll T_{\rm um}$ where only normal processes involving longitudinal phonons are allowed we expect ${\lambda^{\rm ph}_e} \propto v_l^4/T^3$.  However, the normal electron-phonon process is too weak to compete with two other sources of electron scattering that we now discuss. 

\subsection{Electron-impurity scattering}
As we noted in \Sect{composition}, in accreting neutron stars nuclear reactions that process accreted material can produce a mix of metastable nuclei. The evolution of nuclei in the outer crust has been studied in \cite{Gupta:2008} where it was found that electron capture induced neutron emission reactions populate a very diverse mix of nuclei with a large dispersion in $Z$ and $A$.  Although it is reasonable to expect that this dispersion will significantly decrease in the inner crust due to pycno-nuclear reactions and the abundant supply of neutrons, reaction pathways in the inner crust remain poorly understood. 
It is generally assumed that at each depth a specific nucleus with large $Z$ and the highest abundance will crystallize and the remaining mix of nuclei can be treated as impurities in the solid. The impurity parameter 
\begin{equation} 
Q_{\rm imp} = \frac{1}{n_{\rm ion}}~\sum_i~n_i~(Z_i-\langle Z \rangle)^2\,,  
\end{equation}  
is a good measure of the dispersion in the nuclear charge. For moderate $Q_{\rm imp} \approx 1 $ an ordered lattice is likely with scattered impurities. If the impurities cannot diffuse easily their spatial distribution will be uncorrelated, and electron scattering off them can become significant. The scattering mean free path in this case is given by 
\begin{equation}
\lambda_e^{\rm imp} =\frac{k_{\rm Fe}^2}{4 \pi e^4~\sum_i n_i ~(Z_i-\langle \bar{Z }\rangle)^2  }~\Lambda^{-1} =\frac{3 \pi \langle Z \rangle} {4e^4 Q_{\rm imp} ~
k_{\rm Fe}}~\Lambda^{-1}\,,
\end{equation} 
where $\Lambda\simeq 1/2~(\ln{(\pi/e^2)} -2)$ is the Coulomb logarithm, and we have used charge neutrality which requires  $\langle Z \rangle ~n_{\rm ion} = n_e=k_{\rm Fe}^3/3\pi^2$ in arriving at the second equality. 

\subsection{Electron-electron scattering} 
Typically electron-electron scattering is weak but it can become important when electron-ion scattering is suppressed at $T < T_{\rm um}$.   Scattering between relativistic electrons is dominated by the current-current interaction which unlike the Coulomb interaction between charges, this interaction is unscreened in the static limit. The corresponding mean free path was calculated including the effects of dynamical screening (or Landau damping) in \cite{Shternin:2004}.  For the case of relativistic and degenerate electrons 
\be 
\lambda_{\rm e-e} = \frac{\pi^2}{6\zeta[3]~e^2~T} \approx \frac{188}{T} \,, 
\ee  
and it is remarkable that it is independent of density.  The corresponding conductivity $\kappa_{\rm e-e} \simeq 21~\mu_e^2$ is also interesting as it is independent of temperature. Consequently,  electron-electron process can become important at $T < T_{\rm um}$ when electron-phonon Umklapp scattering is suppressed. However, in practice for $T\ge 10^7$ K they are only relevant in a small region close to the crust-core boundary if $Q_{\rm imp} \ll 1$. 

\subsection{Electron conduction}
\begin{figure}[t]
   \centering
   \includegraphics[width=0.95\textwidth]{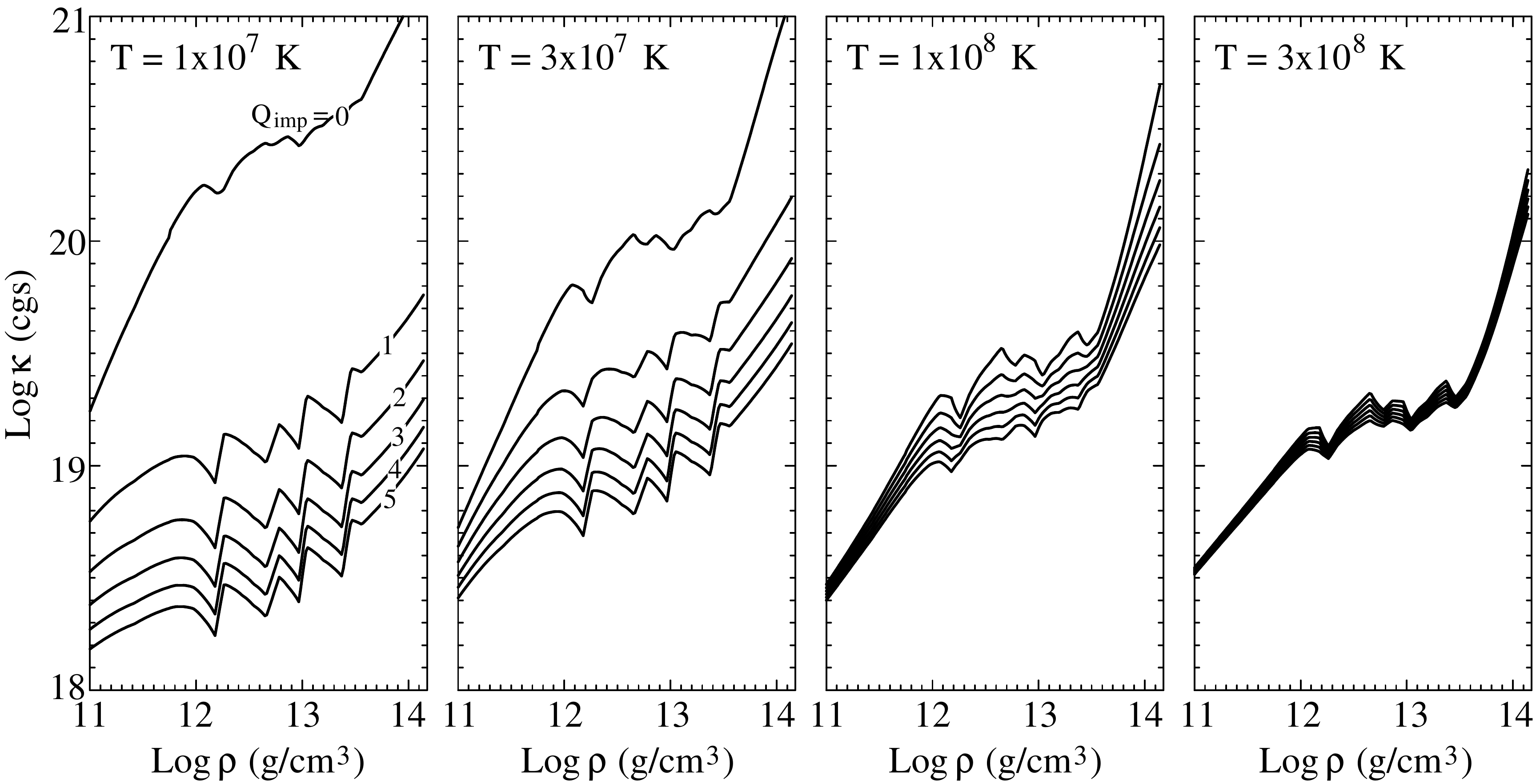} 
   \caption{Electron thermal conductivity $\kappa_e$ vs density at four different temperatures.
   Scattering processes e-ion, e-e, and e-imp with 6 values of $Q_\mathrm{imp} = 0$, 1, 2, 3, 4, and 5
   (as indicated in the left panel) are included.}
   \label{fig:Conductivity}
\end{figure}

Numerical calculations of the electron conductivity with several refinements that include the role of multi-phonon excitations, Debye-Waller corrections and the nuclear form factors have been calculated and tabulated by the neutron star research group at the Ioffe institute in St. Petersburg (http://www.ioffe.rssi.ru/astro/conduct/). Since our focus here is to emphasize the qualitative aspects at low temperature we do not review these important refinements. The results obtained (using the fits to the tabulated results) are shown in Fig.~\ref{fig:Conductivity} and qualitative features can be generally understood in terms of our preceding discussion. Four panels with increasing T in Fig.\ref{fig:Conductivity} clearly demonstrates: (i) the rapid decrease in thermal conductivity for the case $Q_{\rm imp}=0$ as $T$ becomes larger than $T_{\rm um}$ and (ii) the importance of impurity scattering in the inner crust for $T<10^8$ K and for $Q_{\rm imp} \gtrapprox 1$. Both of these trends are easily understood in terms of the preceding discussion of various scattering mechanisms and their temperature dependencies. As we discuss in \Sect{observations}, $Q_{\rm imp}$ will play an important role in interpreting observations in accreting neutron stars when the inner crust is cold with $T < T_{\rm um}$.

\subsection{Phonon conduction} 

Phonon heat conduction can become relevant when $T \gsim 10^8$ K when the phonon heat capacity becomes comparable to that of electrons, or when the electron contribution is suppressed either due to  large $Q_{\rm imp}$ or magnetic fields. Its importnace depends on the phonon mean free path being large enough to compensate for their smaller velocity. Phonon scattering processes have been discussed in Refs.~\cite{Chugunov:2007,Aguilera:2008ed} and we will briefly review them here. As in terrestrial metals \cite{Ziman:1960}, electrons in the inner crust are efficient at damping lattice phonons. The phonon-electron process is shown in  Fig~\ref{fig:Feyn_Diagrams} (2a) which depicts a phonon decay producing an electron-hole excitation. This, Landau damping, dominates over phonon-impurity and phonon-phonon processes for the temperature realized in the crust 
\cite{Chugunov:2007}. 

The electron-phonon process discussed in \Sect{elec_phon} and the phonon-electron process we discuss here are essentially similar. Only here it acts to bring into equilibrium the phonon distribution function that carries the net thermal current relative to the electron gas. Since transverse modes dominate the heat capacity their contribution to thermal conduction is relevant and longitudinal modes can be neglected. For 
$T \ge T_{\rm um}$, Umklapp processes dominate and transverse phonons are absorbed and emitted by large angle electron scattering on the Fermi surface. The mean free path for these processes was estimated by Chugunov and Haensel in \cite{Chugunov:2007}. For simplicity, neglecting corrections due to the Debye-Waller factor, we can rewrite their estimate as  
\begin{eqnarray} 
\lambda^{\rm um}_{\rm lph-e} &=& \frac{\pi~\gamma}{e^2}~\frac{1}{\bar{v}~k_{\rm Fe}}~\frac{F(T_p/T)}{\Lambda_{\rm ph-e}}\approx  \frac{283~a}{ \bar{v}~ Z^{2/3}}~\frac{F(T_p/T)}{\Lambda_{\rm ph-e}}  \,, \\
\label{eq:lambda_lphU}
{\rm where} \nonumber \\
F(T_p/T) &=& 0.014 + \frac{0.03}{\exp{(T_p/5T)}+1} \,,\quad
\Lambda_{\rm ph-e}= \ln \left(\frac{2}{\gamma}\right)-\frac{1}{2}\left(1-\frac{\gamma}{4}\right)
\end{eqnarray}  
where $\bar{v}=\omega_p/3 q_{\rm D}\simeq v_t$ is average phonon velocity and  $\gamma = q_{\rm D}/k_{\rm Fe} = (2/Z)^{1/3}$.  For typical values of $Z\simeq 20-40$ and $\bar{v} \approx 10^{-2}$ the mean free path  $\lambda^{\rm um}_{\rm lph-e} \lessapprox 100~ a$ and the corresponding thermal conductivity is negligible\cite{Chugunov:2007}. Below $T_{\rm um}$ electron Umklapp is suppressed and phonon-electron process cease to operate for transverse phonons. Although their mean free path can be large, numerical calculations show electrons continue to dominate. Normal phonon-electron process will continue to operate for longitudinal phonons at $T< T_{\rm um}$ and their mean free path
\be 
\lambda^{\rm N}_{\rm lph-e} \simeq \frac{\pi}{6~\zeta[3]}~\frac{1}{T}
\ee 
was estimated in \cite{Aguilera:2008ed}. 
\begin{figure}[t]
   \centering
   \includegraphics[width=0.6\textwidth]{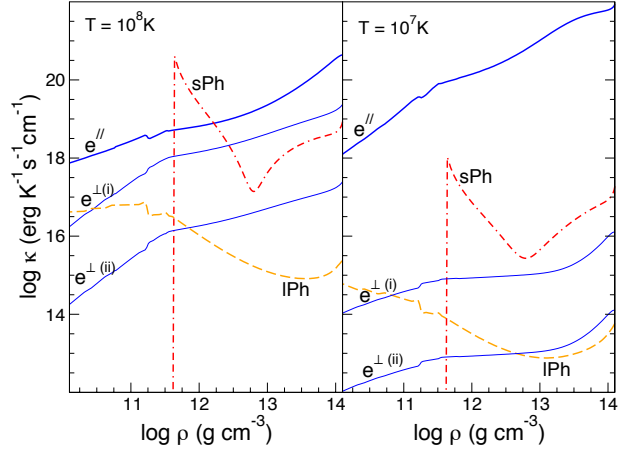} 
   \caption{Electron and phonon thermal conductivity in the presence of large magnetic fields from \cite{Aguilera:2008ed}. }
   \label{fig:phonon_conduction}
\end{figure}

The superfluid can also transport heat. In terrestrial superfluids such a liquid helium extremely efficient heat flow occurs through a non-diffusive process called internal convection. This process which is also responsible for the propagation of second sound in superfluid helium is described as the counter flow of normal and superfluid components in the Landau two-fluid model \cite{Tilley:Book}. In neutron stars the situation was found to be quite different due to strong dissipation of the normal phonon component \cite{Aguilera:2008ed}. Here, heat flow is diffusive as superfluid phonons are effectively scattered by their indirect coupling to electrons. This coupling is induced by the interaction between superfluid and lattice phonons described in \Sect{entrain} and is shown in diagram (3a) of Fig.~\ref{fig:Feyn_Diagrams}. The superfluid phonon mean free path 
\be     
\lambda_{\rm sph-e} = \frac{v^2_\phi}{g_{\rm mix}^2} ~\frac{1+(1-\alpha^2)^2~ (\omega \tau_{\rm lph})^2}{\alpha(\omega \tau_{\rm lph})^2 } ~\lambda_{\rm lph-e}
\label{eq:lambda_sph}
\ee  
where $ \alpha = v_l/v_\phi$, $\tau_{\rm lph}=\lambda_{\rm lph-e}/v_l$ and $\lambda_{\rm lph-e}$ is the mean free path of the longitudinal lattice phonon. Eq.~\ref{eq:lambda_sph} is valid in regions away from the resonance region where $\alpha \approx 1$. Near neutron drip $\alpha \gg 1$ and Eq.~\ref{eq:gmix} can be approximated as  
\be
\lambda_{\rm sph-e} \simeq \frac{v_l^3}{v_\phi^3}~ \frac{n_f (n_b+n_p)}{n_b^2} ~\lambda_{\rm lph-e}=\frac{v_l^3}{v_\phi^3}~ \frac{(A_{\rm cell}-A^*) A^*}{(A^*-Z)^2} ~\lambda_{\rm lph-e}\,, 
\ee 
to deduce that $\lambda_{\rm sph-e} \gg  \lambda_{\rm lph-e}$. Superfluid heat conduction can become relevant here, especially in the presence of large magnetic fields that suppress electron conduction transverse to the magnetic field.  This is illustrated in Fig.~\ref{fig:phonon_conduction}. With increasing density superfluid heat conduction becomes negligible because $\lambda_{\rm sph-e} \approx  \lambda_{\rm lph-e}$ in regions where $ v_l \approx v_\phi$  and when $ v_\phi \gg v_l$ the contribution remain small because $C_v^{\rm phonon} \propto 1/v_\phi^3$ is negligible.

\section{Observable manifestations} 

\lSect{observations}
Possible observational probes of the thermal and transport properties of the inner crust require astrophysical
settings in which the crust evolves rapidly on observationally accessible timescales. 
A first case is simply the early cooling of a new-born neutron star, which relaxes from the initial hot proto-neutron star
stage.
The second family of scenarios involves some process which deposit a large amount of heat in the crust
and observation of the relaxation of the crust can provide invaluable information.
Such heating is expected to occur in a neutron star undergoing accretion from its companion in a binary system
and, in a completely different setting, in magnetar giant flares where catastrophic magnetic realignment deposits a large amount of energy. In both cases thermal relaxation of a heated crust have been observed.
There is also an intriguing possibility that the early evolution of a "neo-neutron star" born in the aftermath of a 
supernovae associated with GRBs has been observed in the late X-ray emission \cite{Negreiros:2011uq}.

\subsection{Thermal evolution equations and crust relaxation}

The equations controlling the time evolution of the neutron star temperature $T$ are
\be
C_V \frac{\partial T}{\partial t} =  Q_h - Q_\nu -\frac{1}{4\pi r^2} \frac{\partial L}{\partial r}
\;\;\;\;\;\;\;\;  \mathrm{and} \;\;\;\;\;\;\;\; 
F =  -\kappa \frac{\partial T}{\partial r}
\;\;\;\;  \mathrm{with} \;\;\;\;
L = 4\pi r^2 F
\label{Eq:Evolution}
\ee
where the first equation simply express energy conservation and the second describes heat transport \footnote{These
are the Newtonian version of the full GR equations, see, e.g., \cite{Page:2004kx}.}.
$F$ and $L$ are the diffusive flux and luminosity, resp., within the star, $C_V$ and $\kappa$ the specific heat and thermal 
conductivity discussed above, $Q_\nu$ is the neutrino emissivity and $Q_h$ the heating rate
(both typically expressed in units of erg cm$^{-3}$ s$^{-1}$). 
Together these two equations give us the heat equation
\be
C_v \frac{\partial T}{\partial t} = \kappa \frac{\partial^2 T}{\partial r^2} + 
\frac{1}{r^2} \frac{\partial (r^2 \kappa)}{\partial r} \frac{\partial T}{\partial r}+ Q_h - Q_\nu
\ee
from which a thermal time-scale can be estimated as
\be
\tau_\mathrm{th} \sim \frac{C_V l^2}{\kappa}
\label{Eq:tau_th}
\ee
where $l$ is a typical length-scale for the $T$ variation.
So, the observation $\tau_\mathrm{th}$ would constrain $C_V$ and $\kappa$ but only through the
combination in  Eq.~\ref{Eq:tau_th}, and modulo an assumed value of $l$.
More information, however, can be obtained in special circumstances which we describe below.

\subsection{Crust Relaxation in Quasi-Persistent Accreting Neutron Stars}

Among neutron stars in accreting binary systems, a very special class called Soft X-Ray Transients (SXRTs) are formed by systems where accretion is transient. Among the 3 dozen known SXRTs, four systems have the peculiarity that the observed accretion outburst(s) lasted several years (while most SXRTs show accretion outbursts lasting a few weeks, separated by months/years/decades of quiescence). These four systems have been dubbed as "Quasi-Persistent Sources" and they are excellent systems to study thermal relaxation. 

In these sources nuclear reactions in the crust deposit heat deep in the crust.  The accreted H/He can be processed to large $A \simeq 100$ by the rapid proton (rp)  capture process \cite{Schatz:2001} and in the event of a superburst this mix  is reprocessed iron-peak nuclei at densities $< 10^{10}$ g cm$^{-3}$ \cite{Schatz:2003}.  Subsequently, when compressed to higher density due to accretion, these nuclei undergo double electron captures
($[A,Z] + 2e \rightarrow [A,Z-2]$) as $\mu_e$ increases,
dripping of neutrons at $\rho > \rho_\mathrm{drip} \simeq 6-7 \times 10^{11}$ g cm$^{-3}$, and, 
when $Z$ has been sufficiently reduced by the electron captures, pycno-nuclear fusions
($[A,Z]+[A,Z] \rightarrow [2A,2Z]$, immediately followed by $[2A,2Z]+2e \rightarrow [2A,2Z-2]$).
This sequence of nuclear reactions injects into the crust, depending on the details of the model, between 1.5-2 MeV
for each accreted nucleon during its journey from the outer crust to the core.
Known as "deep crustal heating" \cite{Brown:1998mi}, it constitutes the $Q_h$ heating term in Eq. \ref{Eq:Evolution}.

In the Quasi-Persistent Sources the accretion outburst last long enough to heat the crust out of its 
initial thermal equilibrium with the core. However, since the core conductivity and heat capacity are very large its temperature 
does not vary appreciably during the outburst. Once accretion stops, the thermal relaxation of the crust can be observed, and is being continuously observed in the four known cases: 
\\

 \noindent {\bf -  KS 1731-260:}
outburst of 12.5 yrs, with an estimated average accretion rate $\langle\dot{M}\rangle \sim 0.1 \dot{M}_\mathrm{Edd}$,
which ended in January 2001 \cite{Cackett:2010bs}; \\
 
\noindent {\bf  - MXB 1659-29:} outburst of 2.5 yrs, $\langle\dot{M}\rangle \sim 0.1 \dot{M}_\mathrm{Edd}$,
which ended in September 2001 \cite{Cackett:2008dz};\\

\noindent{\bf - XTE J1701-462:} outburst of 1.6 yrs, $\langle\dot{M}\rangle \sim \dot{M}_\mathrm{Edd}$,
which ended in August 2007 \cite{Fridriksson:2011fu};\\

\noindent{\bf - EXO 0748-676:} outburst of 24 yrs,  $\langle\dot{M}\rangle \sim 0.01 \dot{M}_\mathrm{Edd}$, 
which ended in August 2008 \cite{Degenaar:2011ij};\\

\noindent where $\dot{M}_\mathrm{Edd} \sim 10^{-8} \, M_\odot$ yr$^{-1}$ $\sim 10^{18}$ g s$^{-1}$ is the Eddington rate.
We briefly present below two case studies.

\subsubsection{Mapping the thermal conductivity: MXB1659-29.}

The crust relaxation of MXB 1659-29 has been studied in detail by Brown \& Cumming in \cite{Brown:2009oq},
and our results amply confirm their analysis.
The accretion outburst was long enough that the crust could reach a steady state:
this is very important since it implies that {\it the initial $T$ profile for the crust relaxation was independent of $C_V$},
providing some relief from the $C_V/\kappa$ degeneracy in $\tau_\mathrm{th}$, Eq. \ref{Eq:tau_th}.

\begin{figure}     
\centering
   \includegraphics[width=0.50\textwidth]{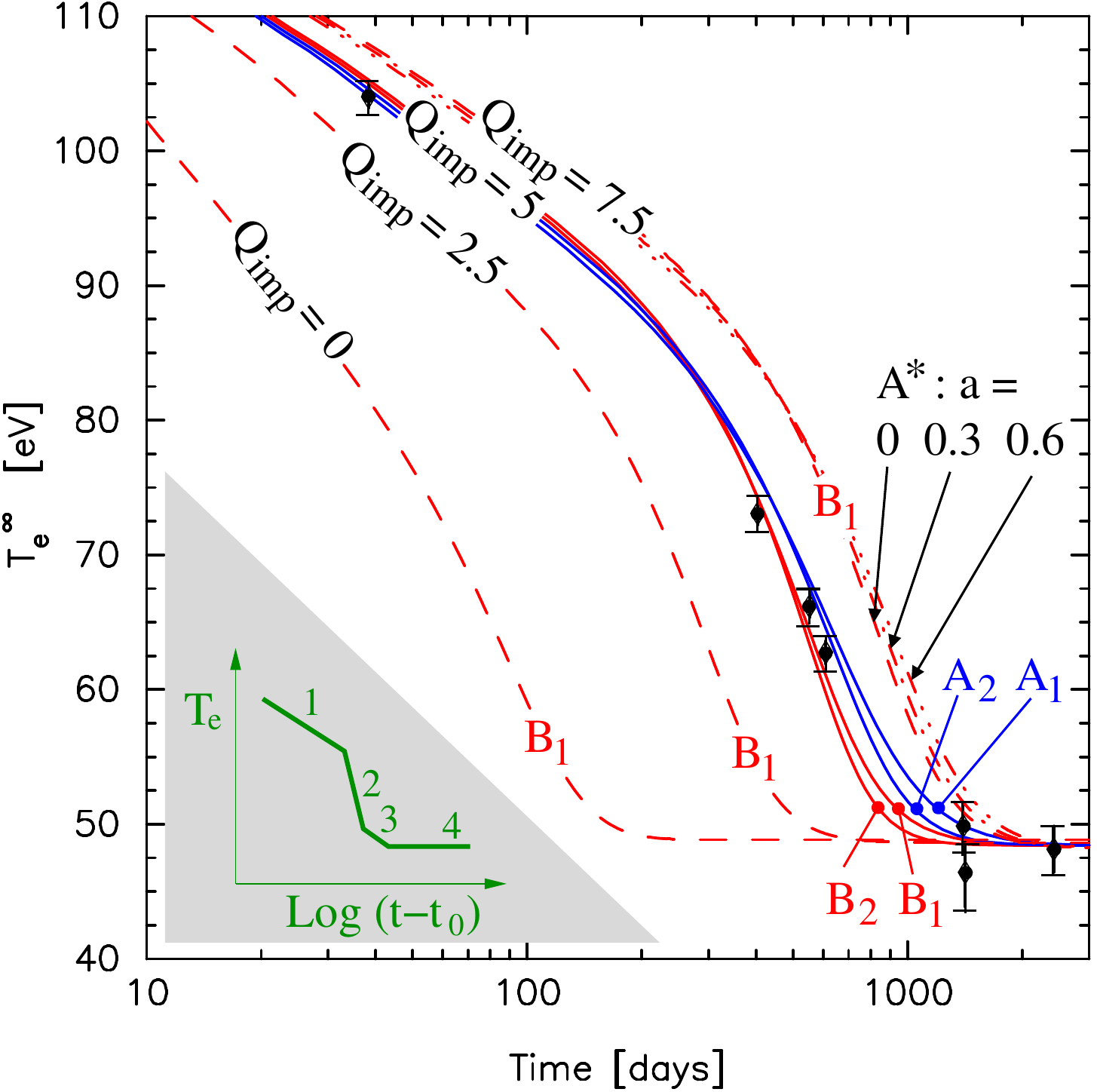} 
      \includegraphics[width=0.49\textwidth]{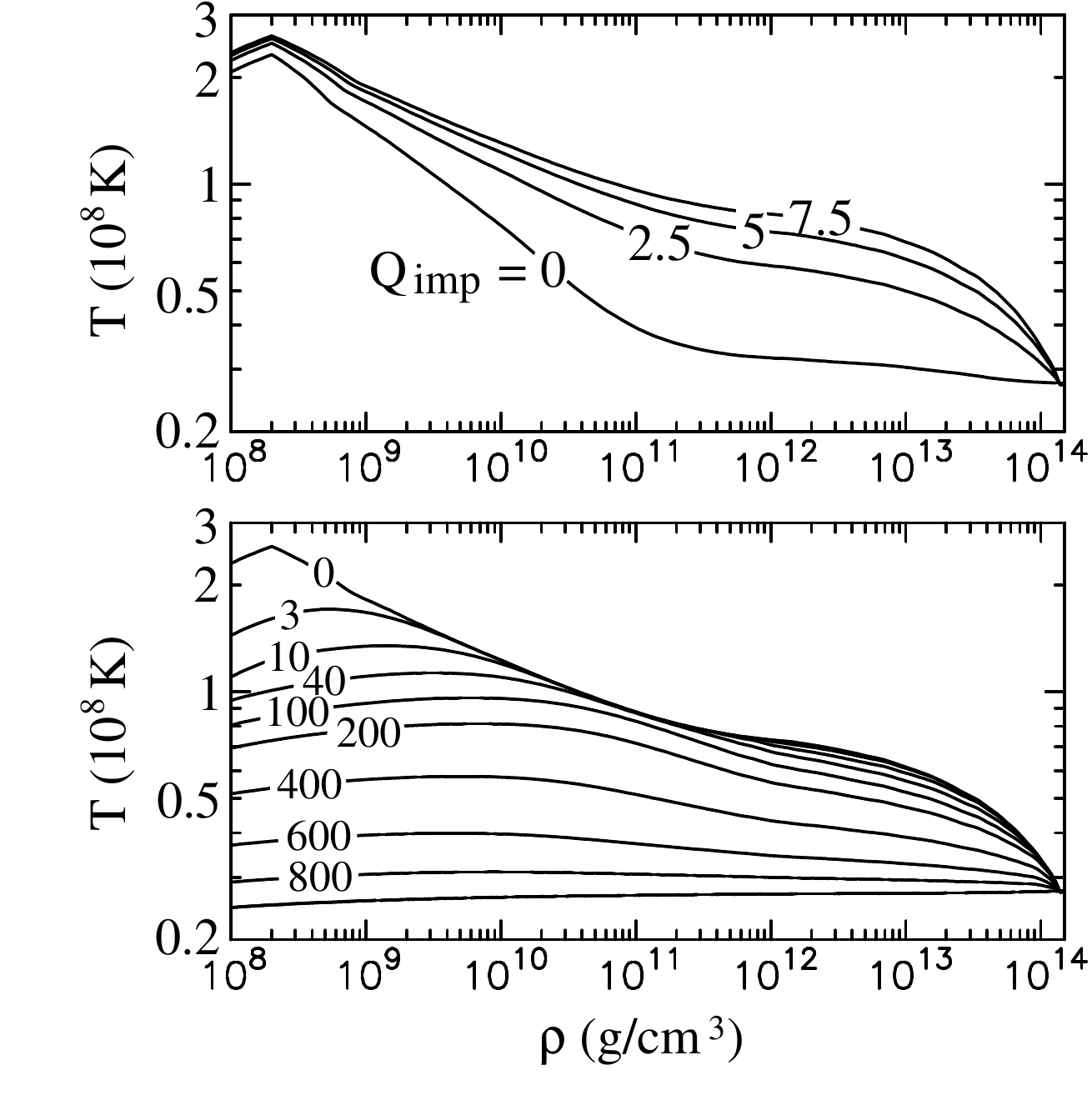}
   \caption{Models for the crust relaxation of MXB1659-29. See text for description.
   The six data points in the left panel are from \cite{Cackett:2008dz}, assuming a source distance of 8.5 kpc.}
   \label{fig:Cool_MXB}
\end{figure}

As was shown in \cite{Brown:2009oq} there is a one-to-one mapping between the cooling curve, $T_e(t)$, and the
temperature profile of the crust, $T(z)$ at the end of the outburst at time $t_0$. At time $t-t_0$ after relaxation commences, the observed surface temperature $T_e$ is determined by the temperature $T(z_{t-t_0})$ at a depth $z_{t-t_0}$ such that the thermal relaxation time  from the surface to this depth is $\tau_\mathrm{th} \sim t $. (This is the "$l^2$-effect" in Eq.\ref{Eq:tau_th}.)
The schematic in the grey shaded inset in the left panel of Fig. \ref{fig:Cool_MXB} shows: phase "1" when $T_e$ is determined by the outer crust evolution;  in "2" it is controlled by the
evolution of matter at densities $\rho \sim 10^{11} - 10^{13}$ g cm$^{-3}$; in phase "3" the evolution is sensitive to the deep inner crust; and, finally, in phase "4", the crust has relaxed with the core and $T_e$ reflects the core temperature.
Approximating $C_V$ and $\kappa$ by power laws in $T$, the evolution is described by power laws, i.e., straight lines
in a $T_e$-Log$(t-t_0)$ plot.

The thermal conductivity of a pure crystalline crust turns out to be much too high to reproduce observed cooling,
but good fits are obtained when $\kappa$ is reduced due to additional scattering by impurities.
The cooling curves in the left panel of Fig. \ref{fig:Cool_MXB} illustrate three cases with impurity parameters
$Q_\mathrm{imp}$ = 2.5, 5, and 7.5, as well as a pure crystalline crust, $Q_\mathrm{imp} =0$.
A value of $Q_\mathrm{imp} = 5$ is favored in this set of results, but is dependent on the assumed crust thickness
and accretion rate (see \cite{Brown:2009oq} for a complete study).
A finer study \cite{Page:kl} with a density dependent $Q_\mathrm{imp}$ reveals that the cooling curves are
mostly sensitive to the value of $Q_\mathrm{imp}$ at $\rho > 10^{13}$ g cm$^{-3}$, so that MXB 1659-29,
and also KS 1731-260, are likely constraining the transport properties of the pasta phase.

In the right panels of Fig. \ref{fig:Cool_MXB} we show samples of $T$ profile in the crust.
In the upper panels are displayed the profiles at the end of the 2.5 yr accretion phase for the four models
with various values of $Q_\mathrm{imp}$ as labeled.
In the lower panel we show the time evolution of the $T$ profile for the model with $Q_\mathrm{imp} = 5$
and the gap model B1 after the end of accretion: times $t-t_0$ in days are indicated on the curves.

An issue that was only briefly discussed in Ref.~\cite{Brown:2009oq} is the range of uncertainty in the specific heat of the inner crust. Either the neutron $C_V^\mathrm{n}$ due to uncertainty the density dependence of  the $^1$S$_0$ superfluid gap or the ion $C_V^\mathrm{I}$ from uncertainty in the effective mass $A^*$ and shear modulus of the ions can alter the evolution timescale.
We illustrate the uncertainty on $C_V^\mathrm{n}$ in the left panel of Fig. \ref{fig:Cool_MXB} in the case $Q_\mathrm{imp} =5$
by displaying four cooling curve for the four neutron $^1$S$_0$ $T_c$ curves of Fig \ref{fig:Tc}.
One sees a divergence of the curves with gaps A1/A2 from the ones with B1/B2 at the beginning of phase 2:
this phase is sensitive to the physics just above the neutron drip point and gaps A's result in a larger 
$C_V^\mathrm{n}$ in this region, resulting in a slower cooling.
During the later phase 3 the cooling curves with the gaps 1's diverge from the ones with gaps 2's:
this phase is sensitive to the physics in the deepest inner crust where the gaps A1 and B1 vanish and
imply a much larger $C_V^\mathrm{n}$ than the gaps A2 and B2.
(The models with $Q_\mathrm{imp} \ne 5$ where all calculated with the gap B1.)
The uncertainty on $C_V^\mathrm{I}$ is illustrated in the models with $Q_\mathrm{imp} =7.5$:
increasing $A^*$, and thus $C_V^\mathrm{I}$, resulting in only moderately slower cooling rates.
(The models with $Q_\mathrm{imp} \ne 7.5$ where all calculated with $A^*=A$.)

These results show that the dominant piece of physics controlling the evolution of MXB1659-29
is the thermal conductivity (and, of course, the crust thickness), and similar conclusions hold for KS 1731-260 \cite{Shternin:2007md}:
{\it crust relaxation of cold neutron stars after a long accretion outburst are perfect laboratories to study the transport
properties of the crust}.

\subsubsection{A peculiar case: XTE 1701-462.}

Crust relaxation after the short, but strong, outburst of  XTE 1701-462 presents distinctive behavior not seen in source with long and less powerful outbursts. XTE 1701-462 is quite orthogonal to MXB1659-29 and KS 1731-260 in that its outburst was shorter, $\dot{M}$ about ten times larger, and its crust about 3 times warmer. It thermal relaxation is fit by a simple exponential with a  decay time of about 100 days, in sharp contradistinction to the 500 days in  MXB1659-29 and KS 1731-260 

\begin{figure}     
\centering
  \includegraphics[width=0.99\textwidth]{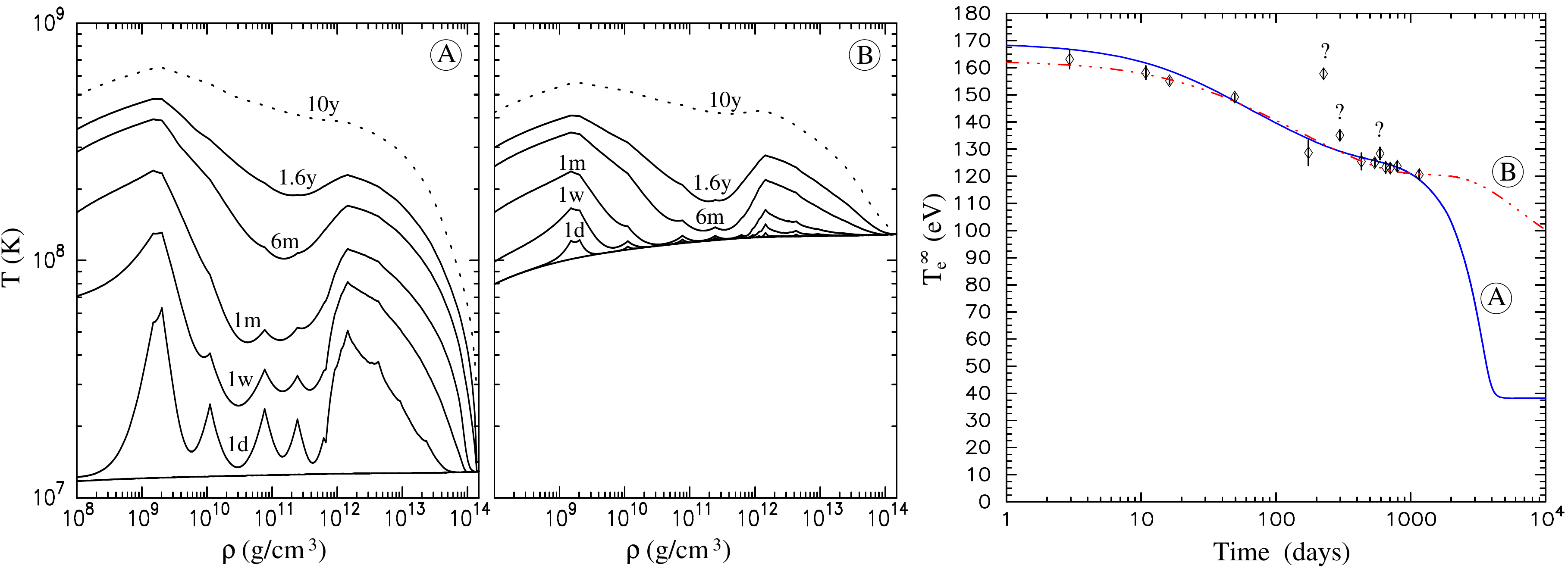}
     \caption{ Models for the crust heating and relaxation of XTE 1701-462
     See text for description.
     The data are from \cite{Fridriksson:2011fu}.}
   \label{Fig:Cool_XTE}
\end{figure}

The system is still cooling and, hence, not enough information is available to draw definitive conclusions,
but some preliminary results are encouraging \cite{Page:2012qa}. Two scenarios that fit well the observations to date are shown in the right panel of Fig. \ref{Fig:Cool_XTE}.  Model "A" having a cold ($\sim 10^7$ K) core and "B" with a hot ($\sim 10^8$ K) core are both able to describe the data reasonably well. We note that the three hot points at days 225, 298, and 593 and marked by a "?" are most likely due to some short phase of residual accretion unrelated to thermal relaxation, (see \cite{Fridriksson:2011fu} for additional details).

The two left panels of Fig. \ref{Fig:Cool_XTE} show the evolution of the $T$ profiles of these two models during the 1.6 yr long accretion outburst. These profiles should be contrasted with the $T$ profile of MXB 1659-29 in the upper right panel of Fig. \ref{fig:Cool_MXB}: it would have taken much more time for XTE 1701-462 to reach a steady $T$ profile.
For comparison, the "10y" profiles in Fig. \ref{Fig:Cool_XTE} correspond to a 10 yrs outburst, instead of 1.6 yr,
and resembles more closely the $T$ profile of MXB 1659-29 at the end of its outburst.
The cusps in the early, 1 day and 1 week, profiles mark the locations of the energy sources:
the major releases occur in the inner crust, from the pycno-nuclear reactions, while below $\rho_\mathrm{drip}$
only electron captures are present. 
The low density region, below $10^{10}$ g cm$^{-3}$, heats up rapidly due to its small specific heat while
the region above it, but below $\rho_\mathrm{drip}$ evolves more slowly due to its larger $C_V$ and
the absence of strong energy sources.
The resulting dip in the $T$ profile below $\rho_\mathrm{drip}$ is precisely what leads to a much more rapid
cooling in the observed light curve, compared to MXB1659-29 and KS 1731-260 where the heating was slower
and the outburst long enough for heat to flow into this dip and produce a much smoother $T$ profile.

Keeping this region below $\rho_\mathrm{drip}$ cold does require a low thermal conductivity, even lower than
what was found in MXB 1659-29 (but as we noted above, this last system constrains $\kappa$ in the high density
pasta phase, but not much in the $\rho_\mathrm{drip}$ region).
From the observation at day 175 till the last observation at day 1158 we see a plateau with a slow cooling:
keeping the crust hot for such a long time also requires a significantly lower conductivity around and below 
$\rho_\mathrm{drip}$.

While the inner crust of MXB1659-29 at the end of the accretion outburst was at 
$T \simless 0.7 \times 10^8$ K, in the case of XTE 1701-462 $T$ reaches $\sim 2 \times 10^8$ K:
in the former case we are in a situation where impurity scattering strongly affected $\kappa$ and
the ions to $C_V$ was small (see Fig.~\ref{fig:Conductivity} and \ref{fig:Cv}).
Thus, not only does the astrophysical setting, high $\dot{M}$ and short outburst, but also the higher $T$ encountered
in XTE 1701-462 make it a peculiar system where the evolution is nearly independent of the electron capacity and electron-impurity  
scattering in the inner crust. The fourth system, EXO 0748-676, will present still another variant: very long outburst, low accretion rate, 
and high $T$ \cite{Page:2012qa}.

\section{Unresolved issues and future directions}
Renewed interest in the neutron star crust has already led to many recent developments and several of these are discussed in this book. In what follows we highlight a few areas where we can anticipate further progress and indicate why these improvements are needed to interpret observations.  
\begin{enumerate} 

\item  \underline {Existence and extent of the pasta phase}: Since a large fraction of the crust (by mass) can be in the pasta phase it is important to determine the surface energy in asymmetric systems, and the density dependence of the nuclear symmetry energy. These are the key inputs needed to calculate the energy of the large lower dimensional structures encountered in the pasta phase. These structures are favored over spherical nuclei only if the surface energy costs are modest.  Calculations of neutron drops and asymmetric matter using realistic nucleon-nucleon interactions and ab-initio methods such as Quantum Monte Carlo have already provided new insights. For example, calculations of the density profiles of neutron drops indicate that Skyrme functionals used in the neutron star context typically underestimate the surface energy \cite{Gandolfi:2011}. An exploration of different density dependencies of the symmetry energy, and the larger isospin gradient energies suggests a much reduced volume of pasta in the crust \cite{Newton:2011dw}. 

\item  \underline {Velocity of shear modes in the inner crust}: As we emphasize in this article at low temperature the velocity of the shear modes of the inner crust is a key input for both thermal and transport properties. The evolution of the shear modulus in the inner crust, especially in the pasta phases can differ greatly from those employed currently. In addition, effects due to entrainment, finite nuclear size and polarizability can all tend to lower the shear speed in the inner crust. A self-consistent study of these effects will be useful to exploring connections between observations of thermal relaxation in a accreting systems and quasi-periodic oscillations in magnetars, since the shear mode plays a crucial role in both phenomena.   

\item \underline{Transport properties of the pasta phase}: At high temperature, $T\gtrapprox1$ MeV, the 
 static \cite{Horowitz:2008vn} and dynamic \cite{Horowitz:2005ff} structure factors of the pasta phase have been calculated using molecular dynamics. However, since the regime of interest in accreting neutron stars is $T \ll T_p < 1$ MeV,  transport properties will depend on the calculation of $S(\omega,k)$ at low temperatures where quantum effects and superfluidity are important.  The shear modulus in the pasta region is still poorly known and will affect the velocity of transverse modes which was shown to play a crucial role in \Sect{thermal} and \Sect{transport}. In addition, other collective excitations unique to lower dimensional pasta structures can be important and have been studied recently in the hydrodynamic limit \cite{di-Gallo:2011fu}. More work is needed to understand how  ({\it al dente}) pasta jiggles in order to calculate the heat capacity and dynamic structure factor at low temperature.

\item \underline{ How and when are Umklapp processes suppressed ?}: The electron band gap is sensitive to short-distance details of the background lattice potential because $k_{\rm Fe} \approx (5-6)/a$. The effect of the nuclear form factor, impurities, and crystal imperfections need to be studied to predict the transition temperature between normal and Umklapp processes in the inner crust. This is especially important when interpreting the differences observed in thermal relaxation of systems with different accretion rates as they sample significantly different temperatures in the inner crust. Since impurity scattering becomes important when $T<T_\mathrm{um}$, observationally inferred values of $Q_\mathrm{imp}$ may be erroneous if Umklapp suppression is incorrectly treated.

\item \underline {Evolution of $Q_{\rm imp}$ in the inner crust}: In accreting neutron stars the reaction paths ways are still not full explored in the inner crust. Here electron capture, pycno-nuclear reactions and  nucleon emission, transfer and capture reaction that determine the abundances are very sensitive to poorly known nuclear structure effects such as pairing and shell gaps in extreme neutron-rich nuclei. It remains to be seen if generic expectations for $Q_{\rm imp}$ derived from nuclear structure studies are compatible with the inference that  $Q_{\rm imp}\approx 5$ from the analysis of thermal relaxation observed in MXB 1659-29.

\item \underline{ Are nuclear excitations  relevant for T ~$10^8 -10^9$ K ?}: Nuclear excitations are typically ignored in the crust for $T \lsim 10^{10}$ K. However, low-lying pygmy resonances, related collective surface modes, and perhaps even single particle excitations in the extreme nuclei encountered may affect both the thermodynamics and transport properties if they occur at low energy $\ll \omega_p$ . 
\end{enumerate} 

\section*{Acknowledgments}

D.P.'s work is supported by grants from Conacyt (CB-2009-01, \#132400) and UNAM-DGAPA (PAPIIT, \#IN113211).
The work of S.R. was supported by the DOE grant \#DE-FG02-00ER41132 and by the 
Topical Collaboration to study {\it Neutrinos and nucleosynthesis in hot dense matter}. 
Ed Brown, Vincenzo Cirigliano, Nicholas Chamel, Andrew Cumming, Chris Pethick, and Rishi Sharma are
acknowledged for useful discussions and or collaborations.

\label{lastpage-01}


\begin{thebibliography}{10}

\bibitem{Chamel:2008LRR}
{Chamel}, N. and {Haensel}, P. 
{\em Living Reviews in Relativity} {\bf 11}, 10 (2008)

\bibitem{Baym:1971pi}
{Baym}, G., {Bethe}, H.~A., and {Pethick}, C.~J. 
{\em Nucl. Phys. A} {\bf 175}, 225 (1971)

\bibitem{Negele:1973ve}
{Negele}, J.~W. and {Vautherin}, D.
{\em Nucl. Phys. A} {\bf 207}, 298 (1973)

\bibitem{Pethick:1995}
{Pethick}, C.~J. and {Ravenhall}, D.~G.
{\em Annu. Rev. Nucl. Part. Sci.} {\bf 45}, 429 (1995)

\bibitem{Gupta:2008}
{Gupta}, S.~S., {Kawano}, T., and {M{\"o}ller}, P.
{\em Phys. Rev. Lett.} {\bf 101(23)}, 231101 (2008)

\bibitem{Haensel:2008ly}
{Haensel}, P. and {Zdunik}, J.~L. 
{\em Astron. Astrophys.} {\bf 480}, 459 (2008)

\bibitem{Yakovlev:2006}
{Yakovlev}, D.~G., {Gasques}, L., and {Wiescher}, M. 
{\em Mon. Not. R. Astron. Soc.} {\bf 371}, 1322 (2006)

\bibitem{Kittel:1976}
Kittel, C. (1976)
Introduction to solid state physics,
Wiley, New York ,  5th ed. edition.

\bibitem{RaikhYakovlev_82}
{Raikh}, M.~E. and {Yakovlev}, D.~G. 
{\em Astrophys. Sp. Sci.} {\bf 87}, 193 (1982)

\bibitem{Ginzburg:1969}
Ginzburg, V.~L.
{\em J. Stat. Phys.} {\bf 1}, 3 (1969)

\bibitem{Gorkov:1961kx}
{Gorkov}, L.~P. and {Melik-Barkhudarov}, T.~K.
{\em Sov. Phys. JETP} {\bf 13}, 1018 (1961)

\bibitem{Chen:1993ys}
{Chen}, J.~M.~C., {Clark}, J.~W., {Dav{\'e}}, R.~D., and {Khodel}, V.~V.
{\em Nucl. Phys. A} {\bf 555}, 59 (1993)

\bibitem{Wambach:1993kl}
{Wambach}, J., {Ainsworth}, T.~L., and {Pines}, D.
{\em Nucl. Phys. A} {\bf 555}, 128 (1993)

\bibitem{Gezerlis:2008tg}
{Gezerlis}, A. and {Carlson}, J.
{\em Phys. Rev. C} {\bf 77(3)}, 032801 (2008)

\bibitem{Gandolfi:2008hc}
{Gandolfi}, S., {Illarionov}, A.~Y., {Fantoni}, S., {Pederiva}, F., and  {Schmidt}, K.~E.
{\em Phys. Rev. Lett.} {\bf 101(13)}, 132501 (2008)

\bibitem{Gezerlis:2011uq}
{Gezerlis}, A. and {Carlson}, J.
{\em ArXiv:} 1109.4946 (2011)

\bibitem{1992Natur.360...48C}
{Chabrier}, G., {Ashcroft}, N.~W., and {Dewitt}, H.~E. 
{\em Nature} {\bf 360}, 48 (1992)

\bibitem{Baiko:2001qf}
{Baiko}, D.~A., {Potekhin}, A.~Y., and {Yakovlev}, D.~G.
{\em Phys. Rev. E} {\bf 64(5)}, 057402 (2001)

\bibitem{Margueron:2012ph} 
 Margueron, J. and Sandulescu, N.
  arXiv:1201.2774 [nucl-th].

\bibitem{Chamel:2008ju}
Chamel, N., Margueron, J., and Khan, E.
{\em Phys. Rev. C} {\bf 79}, 012801 (2009)

\bibitem{Cirigliano:2011}
{Cirigliano}, V., {Reddy}, S., and {Sharma}, R.
{\em Phys. Rev. C} {\bf 84}, 045809 (2011)

\bibitem{Carter:2004pp}
Carter, B., Chamel, N., and Haensel, P.
{\em Nucl. Phys. A} {\bf 748(3-4)}, 675 (2005)

\bibitem{Carter:2006}
{Carter}, B., {Chamel}, N., and {Haensel}, P.
{\em Int. J. of Mod. Phys. D} {\bf 15}, 777 (2006)

\bibitem{Chamel:2005}
{Chamel}, N.
{\em Nucl. Phys. A} {\bf 747}, 109 (2005)

\bibitem{Pethick:2010}
{Pethick}, C.~J., {Chamel}, N., and {Reddy}, S.
{\em Prog. Theor. Phys. Supp.} {\bf 186}, 9 (2010)

\bibitem{Chugunov:2007}
Chugunov, A. and Haensel, P.
{\em Mon. Not. R. Astron. Soc.} {\bf 381}, 1143 (2007)

\bibitem{Aguilera:2008ed}
Aguilera, D.~N., Cirigliano, V., Pons, J.~A., Reddy, S., and Sharma, R.
{\em Phys. Rev. Lett.} {\bf 102(9)}, 091101 (2009)

\bibitem{FlowersItoh:1976}
{Flowers}, E. and {Itoh}, N. May
{\em Astrophys. J.} {\bf 206}, 218 (1976)

\bibitem{Ziman:1960}
Ziman, J. (1960)
{Electrons and Phonons},
Cambridge University Press, London, England.

\bibitem{FetterWalecka:2003}
Fetter, A.~L. and Walecka, J.~D. (2003)
Quantum Theory of Many-Particle Systems,
Dover.

\bibitem{YakovlevUrpin:1980}
{Yakovlev}, D.~G. and {Urpin}, V.~A.
{\em Sov. Astron.} {\bf 24}, 303 (1980)

\bibitem{Shternin:2004}
{Shternin}, P.~S. and {Yakovlev}, D.~G. 
{\em Phys. Rev. D} {\bf 74(4)}, 043004 (2006)

\bibitem{Tilley:Book}
Tilley, D. and Tilley, J. (1990)
Superfluidity and Superconductivity,
IOP Publishing Ltd., Bristol.

\bibitem{Negreiros:2011uq}
{Negreiros}, R., {Ruffini}, R., {Bianco}, C.~L., and {Rueda}, J.~A.
 {\em ArXiv:} 1112.3462 (2011)

\bibitem{Page:2004kx}
{Page}, D., {Lattimer}, J.~M., {Prakash}, M., and {Steiner}, A.~W.
{\em Astrophys. J. Supp.} {\bf 155}, 623 (2004)

\bibitem{Schatz:2001}
{Schatz}, H., {Aprahamian}, A., {Barnard}, V., {Bildsten}, L., {Cumming}, A.,
  {Ouellette}, M., {Rauscher}, T., {Thielemann}, F.-K., and {Wiescher}, M.
{\em Phys. Rev. Lett.} {\bf 86}, 3471 (2001)

\bibitem{Schatz:2003}
{Schatz}, H., {Bildsten}, L., and {Cumming}, A.
{\em Astrophys. J. Lett.} {\bf 583}, 87 (2003)

\bibitem{Brown:1998mi}
{Brown}, E.~F., {Bildsten}, L., and {Rutledge}, R.~E.
{\em Astrophys. J. Lett.} {\bf 504}, 95 (1998)

\bibitem{Cackett:2010bs}
{Cackett}, E.~M., {Brown}, E.~F., {Cumming}, A., {Degenaar}, N., {Miller},
  J.~M., and {Wijnands}, R.
{\em Astrophys. J. Lett.} {\bf 722}, 137 (2010)

\bibitem{Cackett:2008dz}
{Cackett}, E.~M., {Wijnands}, R., {Miller}, J.~M., {Brown}, E.~F., and
  {Degenaar}, N.
{\em Astrophys. J. Lett.} {\bf 687}, 87 (2008)

\bibitem{Fridriksson:2011fu}
{Fridriksson}, J.~K., {Homan}, J., {Wijnands}, R., {Cackett}, E.~M.,
  {Altamirano}, D., {Degenaar}, N., {Brown}, E.~F., {M{\'e}ndez}, M., and
  {Belloni}, T.~M.
{\em Astrophys. J.} {\bf 736}, 162 (2011)

\bibitem{Degenaar:2011ij}
{Degenaar}, N., {Wolff}, M.~T., {Ray}, P.~S., {Wood}, K.~S., {Homan}, J.,
  {Lewin}, W.~H.~G., {Jonker}, P.~G., {Cackett}, E.~M., {Miller}, J.~M.,
  {Brown}, E.~F., and {Wijnands}, R.
{\em Mon. Not. R. Astron. Soc.} {\bf 412}, 1409 (2011)

\bibitem{Brown:2009oq}
{Brown}, E.~F. and {Cumming}, A.
{\em Astrophys. J.} {\bf 698}, 1020 (2009)

\bibitem{Shternin:2007md} 
Shternin, P.~S., Yakovlev, D.~G., Haensel, P., and Potekhin, A.~Y.
{\em Mon. Not. R. Astron. Soc.} {\bf 382}, L43 (2007)

\bibitem{Page:kl}
{Page}, D. and {Reddy}, S.
Work in progress.

\bibitem{Page:2012qa}
{Page}, D. and {Reddy}, S.
In preparation (2012).

\bibitem{Gandolfi:2011}
{Gandolfi}, S., {Carlson}, J., and {Pieper}, S.~C.
{\em Phys. Rev. Lett.} {\bf 106(1)}, 012501 (2011)

\bibitem{Newton:2011dw}
Newton, W., Gearheart, M., and Li, B.-A. 
{\em ArXiv:} 1110.4043 (2011).

\bibitem{Horowitz:2008vn}
{Horowitz}, C.~J. and {Berry}, D.~K. 
{\em Phys. Rev. C} {\bf 78(3)}, 035806 (2008)

\bibitem{Horowitz:2005ff}
{Horowitz}, C.~J., {P{\'e}rez-Garc{\'{\i}}a}, M.~A., {Berry}, D.~K., and
  {Piekarewicz}, J.
{\em Phys. Rev. C} {\bf 72(3)}, 035801 (2005)

\bibitem{di-Gallo:2011fu}
{di Gallo}, L., {Oertel}, M., and {Urban}, M.
{\em Phys. Rev. C} {\bf 84(4)}, 045801 (2011)

\end{thebibliography}
\end{document}